\documentclass{aa}
\usepackage{psfig}
\def\cm2{\mbox{$\mbox{cm}^{-2}$}}
\def\cm3{\mbox{$\mbox{cm}^{-3}$}} 
\def\sm{\mbox{M$_\odot$}}

\def\l13{\mbox{$L_{\mbox{\tiny 1.3mm}}$}}
\def\k13{\mbox{$\kappa_{\mbox{\tiny 1.3mm}}$}}

\def\f800{\mbox{$F_{\mbox{\tiny 800}}$}}

\def\>{$>$}
\def\<{$<$}

\def\ltsima{$\; \buildrel < \over \sim \;$}
\def\simlt{\lower.5ex\hbox{\ltsima}}
\def\gtsima{$\; \buildrel > \over \sim \;$}
\def\simgt{\lower.5ex\hbox{\gtsima}}

\begin{document}

   \thesaurus{           
              (08.05.1;  
               09.08.1)  
             }
   \title{An ISOCAM absorption survey of the structure of pre-stellar cloud cores\thanks{Based on observations with ISO, an ESA project with instruments funded by ESA Member States (especially the PI countries: France, Germany, the Netherlands and the United Kingdom) and with the participation of ISAS and NASA.}}

   \author{A. Bacmann$^{1}$\thanks{Present address: Astrophysikalisches Institut und Universit\"ats Sternwarte, Schillerg\"a\ss{}chen 2-3, D-07745 Jena, Germany, e-mail: bacmann@astro.uni-jena.de}, 
P. Andr\'e$^{1}$\thanks{e-mail: pandre@cea.fr}, 
J.-L. Puget$^{2}$, A. Abergel$^{2}$, S. Bontemps$^{3}$, 
D. Ward-Thompson$^{4}$}

   \offprints{A. Bacmann}

   \institute{$^1$CEA, DSM, DAPNIA, Service d'Astrophysique, 
C.E. Saclay, F-91191 Gif-sur-Yvette Cedex, France \\
$^2$Institut d'Astrophysique Spatiale, Orsay, France \\
$^3$Observatoire de Bordeaux, Floirac, France \\
$^4$Department of Physics \& Astronomy, Cardiff University, PO Box 913, Cardiff, U.K.}

   \date{Received 28 October 1999 / accepted 8 March 2000}

   \maketitle
\markboth { Bacmann et al.: An ISOCAM absorption survey of the structure of pre-stellar cloud cores}{ }

   \begin{abstract}

        We present the results of a mid-infrared ($\lambda\simeq$~7~$\mu$m) imaging survey of a sample 
of 24 starless dense cores carried out at an angular resolution of 6$\arcsec$ with the ISOCAM camera aboard the 
Infrared Space Observatory (ISO). The targeted cores are believed to be 
pre-stellar in nature and to represent the initial conditions 
of low-mass, isolated star formation.  
In previous submillimeter dust continuum studies of such pre-stellar cores, 
it was found that the derived column density profiles did not follow a single power-law such as $N_{H_{2}} \propto \bar{r}^{-1}$ throughout their full extent but flattened out near their center. 
These submillimeter observations however could not constrain the 
density profiles at radii greater than $\sim$ 10000~AU. The present absorption study 
uses ISOCAM's sensitivity to map these pre-stellar cores in absorption
against the diffuse mid-infrared background. The goal was to determine their structure
 at radii that extend beyond the limits of sensitivity of the submillimeter continuum maps and at twice
 as good an angular resolution. Among the 24 cores observed in our 
survey, a majority of them show deep absorption features. 
The starless cores studied here all show a column density profile that 
flattens in the center, which confirms the submillimeter emission results. Moreover, beyond a radius of $\sim 5000-10000$ AU, 
the typical column density profile steepens with distance from core center 
and gets steeper than $N_{H_{2}} \propto \bar{r}^{-1}$, until it eventually merges with the low-density ambient molecular cloud. 
At least three of the cores 
present sharp edges at $R \sim 15000-30000$~AU and appear to be decoupled 
from their parent clouds, providing finite reservoirs of mass for subsequent 
star formation.

   \end{abstract}
  

\section{Introduction}

Although our general understanding of low-mass star formation has 
made significant 
theoretical and observational progress in the past three decades 
(see, e.g., the volumes by Levy \& Lunine 1993 and 
Mannings, Boss, \& Russell 2000 for comprehensive reviews), 
the earliest stages of the star formation process remain poorly known. 
These stages immediately preceding and immediately following the onset 
of protostellar collapse are of great interest since to some 
extent they must govern the origin of stellar masses, i.e., 
the stellar initial mass function.

According to present ideas on isolated star 
formation (e.g. Shu, Adams, \& Lizano 1987), the first evolutionary phase  
in the path from molecular cloud to main sequence star involves the formation 
of gravitationally-bound, starless dense cores (e.g. Myers 1999). 
These ``pre-protostellar'' (or ``pre-stellar'' for brevity) cores (cf. Ward-Thompson et al. 1994) are thought to be initially supported 
against their self-gravity by  
magnetic and/or turbulent pressure, and to progressively evolve toward higher
degrees of central concentration through ambipolar diffusion (e.g. Mouschovias
1991) and/or the dissipation of turbulence (e.g. Nakano 1998).
At some not yet well understood point, the cores become unstable
and collapse dynamically to form accreting protostars which quickly become detectable 
in the radio range as Class~0 objects (Andr\'e, Ward-Thompson, \& Barsony 1993) and subsequently in the infrared as Class~I sources (eg. Lada 1987). 
Additional complications arise in regions of multiple star formation (such
as the $\rho$ Ophiuchi central cloud) where external triggers rather than
self-initiated ambipolar diffusion may be responsible for cloud
fragmentation and core formation (e.g. Loren \& Wootten 1986; Boss 1995;
Whitworth et al. 1996; 
Motte, Andr\'e, \& Neri 1998 -- hereafter MAN98).

A good knowledge of the initial conditions for rapid protostellar collapse 
in star-forming cores is crucial to get at a proper theoretical 
description of early protostar evolution. 
In particular, recent studies show that the mass-infall rate during the 
protostellar phase is quite sensitive to the core radial density profile 
at the onset of dynamical collapse.  
When the initial density profile is flatter than $\rho \propto r^{-2} $ near  
core center, 
then the mass-infall rate is expected to feature a strong initial peak, 
followed by a sharp decline, at the beginning of the main accretion phase 
(e.g. Foster \& Chevalier 1993; Henriksen, Andr\'e, \& Bontemps 1997). 
The outer core density profile 
also plays a fundamental role by determining whether the mass 
reservoir ultimately available for star formation is effectively 
finite or infinite. 
If the radial density profile approaches $\rho \propto r^{-2} $ 
up to large radii 
as in the standard paradigm of isolated star formation 
(Shu et al. 1987), then the mass reservoir is much larger than a typical 
stellar mass and feedback processes such as outflow/inflow interactions  
(e.g. Velusamy \& Langer 1998) must eventually stop accretion onto 
the central protostar. On the other hand, if the initial core density 
profile becomes steeper than $\rho \propto r^{-3}$ beyond some finite radius, 
then the mass reservoir is limited and 
the mass accretion rate will naturally drop to negligible values after 
some time. In this case, the final stellar mass may be essentially 
determined at the pre-collapse stage.

Direct observational constraints on the radial density structure at the 
onset of protostellar collapse have been obtained through (sub)millimeter 
dust continuum mapping of a large number of starless cores and 
Class 0/Class I protostellar envelopes  
with the JCMT and IRAM~30~m telescopes 
(e.g. Ward-Thompson et al. 1994; Andr\'e, Ward-Thompson, 
\& Motte 1996 -- hereafter AWM96; Ward-Thompson, Motte, \& Andr\'e 1999 -- hereafter
WMA99). 
General trends have been found: while protostellar envelopes 
are always highly centrally condensed, with radial density profiles 
consistent with $\rho \propto r^{-1.5}$ or $\rho \propto r^{-2}$ 
(e.g. Ladd et al. 1991, Motte \& Andr\'e 1999), starless cores have 
radial profiles that 
flatten out near their centers and become 
much flatter than $\rho \propto r^{-2}$ at radii less than a few thousand AUs
(see Andr\'e, Ward-Thompson, \& Barsony 2000 for a review).
%
It is however necessary to bear in mind the limitations of (sub)millimeter 
continuum emission maps: since the dust {\it emission} from the outer parts of 
cores is intrinsically very weak, (sub)millimeter mapping is 
mostly sensitive to the inner density structure of starless cores. 
The outer density structure of cores and the possible presence of edges 
are difficult to constrain from 
(sub)millimeter observations. Fortunately, starless cores can also be studied 
in {\it absorption} at mid-IR wavelengths (e.g. Abergel et al. 1996, 
Egan et al. 1998), and this technique turns out to be more sensitive to 
the outer parts of cores. Another advantage of the mid-IR absorption 
technique over (sub)millimeter emission mapping  
is that the column density structure of the absorbing material 
may be derived without any assumption about the dust temperature distribution.
 
 

In order to gain further, independent insight into cloud core structure 
prior to protostar formation, we undertook 
dedicated mid-IR imaging observations toward a broad sample of starless 
dense cores with the ISOCAM mid-IR camera aboard the 
{\it Infrared Space Observatory} (ISO). The present paper discusses 
the results of this mid-IR imaging program, whose 
immediate goal was to detect the sample cores 
in absorption against the diffuse mid-IR background produced by 
the mean galactic interstellar radiation field (e.g. Mathis, Mezger \& Panagia 1983) and/or arising from the envelopes of the 
parent molecular clouds (e.g. Bernard et al. 1993).

The layout of the paper is as follows. In Sect.~2, we give the 
properties of the core sample and explain observational details. In Sect.~3, 
we show our ISOCAM absorption maps and analyze the associated absorption 
profiles in terms of column density structure. In Sect.~4, we compare 
our results with various theoretical models of core structure and discuss 
possible implications for our understanding of the pre-stellar stage of 
star formation. Our main conclusions are summarized in Sect.~5.
 
\section{Observations and Data Reduction}

\subsection{Starless core sample}

The 24 starless fields of the present survey were selected from various 
dark cloud and nebula catalogs (e.g. Parker 1988, Benson \& Myers 1989, 
Bourke, Hyland \& Robinson 
1995, Schneider \& Elmegreen 1979), on the grounds that they contained no IRAS
point sources associated with a young stellar object and had an estimated mid-infrared (mid-IR) background stronger than $\sim
1$ MJy/sr at $\sim$~7~$\mu$m. We also included a map of the dark cloud Oph~D (also called L1696A)
situated in the $\rho$ Ophiuchi complex, and already observed in absorption by
ISOCAM (Abergel et al. 1998). The background at 7~$\mu$m was originally estimated from IRAS
images, as were the 7~$\mu$m flux densities of the IRAS sources appearing in 
the fields. Fields containing IRAS point sources with estimated flux densities larger than 1.5~Jy 
at 7.75~$\mu$m (LW6 filter [7-8.5~$\mu$m]) or larger than 0.7~Jy at 6.75~$\mu$m 
(LW2 filter [5-8.5~$\mu$m]) had to be shifted away from the sources and/or the integration 
time had to be shortened in order to avoid saturating 
the ISOCAM detectors (see 2.2 below). 
 The selected regions 
span a wide range of properties (e.g. morphologies, environments), but are all 
located in nearby molecular cloud complexes (distance $\leq$ 500 pc) either 
actively forming stars or supposed to be future 
hosts of star-formation activity.  

\begin{table*}[htpb]
\caption[]{List of starless cores observed with ISOCAM}
\begin{minipage}{16cm}
\begin{flushleft}
\begin{tabular}{lccccccccl} \hline
Source & R.A.(2000) & Dec. (2000) & Dist. & Filter & Map RMS & $<$I$_{MIR}>$ & I$_{zodi}^{\ddagger}$ & Absorp. & References \\
       &            &             &          &        &             &          &                &  Contrast    &        \\
       &            &             & (pc)     &        &    (MJy/sr) & (MJy/sr) &    (MJy/sr)    &      (\%)    &        \\
\hline \hline
L1517B$^{*}$ & 04$^h$55$^m$05$\fs$2 & 30$\degr$38$\arcmin$39$\arcsec$ & 140& LW2 & 0.07 & 4.4 & 3.8 & 5 & (4),(6),(13)\\
L1512$^{*}$ & 05$^h$04$^m$09$\fs$7 & 32$\degr$43$\arcmin$09$\arcsec$ & 140 & LW6 & 0.15 & 12.4 & - & 3 & (4),(5),(6),(11),(15)\\
L1544$^{*}$ & 05$^h$04$^m$18$\fs$1 & 25$\degr$11$\arcmin$08$\arcsec$ & 140  & LW2 & 0.05 & 4.8 & 4.2 & 8 & (4),(11),(14),(15),(16)\\
L1582A & 05$^h$32$^m$03$\fs$0 & 12$\degr$30$\arcmin$16$\arcsec$ & 400 & LW2 & 0.16 & 8.6 & 3.7 & 44 & (4),(6)\\
L1672 & 05$^h$54$^m$24$\fs$0 & 01$\degr$58$\arcmin$53$\arcsec$ & 200 & LW2 & 0.19 & 4.7 & 3.4 & 31 & (10)\\
BHR78 & 12$^h$36$^m$18$\fs$7 & $-$63$\degr$12$\arcmin$37$\arcsec$ & 200 & LW6 & 0.61 & 17.2 & - & 11 & (3)\\
BHR111 & 15$^h$42$^m$19$\fs$7 & $-$52$\degr$47$\arcmin$54$\arcsec$ & 250 & LW6 & 0.41 & 19.5 & 10.0 & 6 & (3)\\
R7$^{**}$ & 16$^h$21$^m$45$\fs$9 & $-$23$\degr$42$\arcmin$51$\arcsec$ & 160 & LW6 & 0.92 & - & - & - & (8)\\
OphD$^{**}$ & 16$^h$28$^m$30$\fs$4 & $-$24$\degr$18$\arcmin$29$\arcsec$ & 160 & LW2$^{\dagger}$ & 0.3 & 14.8 & 5.7 & 28 & (1),(8),(9)\\
R53$^{**}$ & 16$^h$31$^m$44$\fs$1 & $-$24$\degr$48$\arcmin$12$\arcsec$ & 160 & LW2 & 0.2 & 11.4 & - & 20 & (8)\\
SA187 & 16$^h$32$^m$17$\fs$2 & $-$44$\degr$53$\arcmin$40$\arcsec$ & 200 & LW6 & 0.92 & - & - & 6 & (12)\\
L1709A$^{**}$ & 16$^h$32$^m$42$\fs$1 & $-$23$\degr$54$\arcmin$09$\arcsec$ & 160 & LW2 & 0.17 & 12.8 & 10.5 & 17 & (4),(6),(8),(11)\\
R60$^{**}$ & 16$^h$33$^m$11$\fs$8 & $-$24$\degr$41$\arcmin$30$\arcsec$ & 160 & LW2 & 0.13 & 10.1 & - & 7 & (8)\\
L1709C$^{**}$ & 16$^h$33$^m$53$\fs$4 & $-$23$\degr$42$\arcmin$32$\arcsec$ & 160 & LW2 & 0.10 & 8.2 & - & 16 & (4),(6),(8),(11)\\
R63$^{**}$ & 16$^h$33$^m$58$\fs$3 & $-$24$\degr$30$\arcmin$03$\arcsec$ & 160 & LW2 & 0.14 & 9.3 & 5.0 & 16 & (8)\\
L1689B$^{**}$ & 16$^h$34$^m$40$\fs$1 & $-$24$\degr$37$\arcmin$00$\arcsec$ & 160 & LW2 & 0.05 & 8.3 & 4.3 & 19 & (2),(4),(8),(11),(15)\\
L204B & 16$^h$47$^m$32$\fs$0 & $-$11$\degr$59$\arcmin$15$\arcsec$ & 170 & LW2 & 0.11 & 15.7 & - & 4 & (4),(11)\\
B68 & 17$^h$22$^m$34$\fs$7 & $-$23$\degr$47$\arcmin$50$\arcsec$ & 200 & LW2 & 0.19 & 11.5 &  8.8 & 2 & (4),(10)\\
BHR140 & 17$^h$22$^m$53$\fs$9 & $-$43$\degr$22$\arcmin$13$\arcsec$ & 400 & LW6 & 0.34 & 10.2 & 6.8 & 9 & (3)\\
L310 & 18$^h$07$^m$11$\fs$9 & $-$18$\degr$21$\arcmin$35$\arcsec$ & 200 & LW6 & 0.54 & 30.0 & 7.1 & 39 & (11)\\
L328 & 18$^h$17$^m$00$\fs$4 & $-$18$\degr$01$\arcmin$52$\arcsec$ & 200 & LW6 & 0.7 & 42.9 & 7.4 & 40 & (11)\\
L429 & 18$^h$17$^m$06$\fs$4 & $-$08$\degr$15$\arcmin$51$\arcsec$ & 200 & LW6 & 0.2 & 8.6 & 6.8 & 17 & (11)\\
GF5 & 18$^h$39$^m$16$\fs$4 & $-$06$\degr$38$\arcmin$15$\arcsec$ & 200 & LW6 & 2.0 & 75.2 & 7.2 & 30 & (13)\\
B133 & 19$^h$09$^m$08$\fs$8 & $-$06$\degr$53$\arcmin$20$\arcsec$ & 400 & LW6 & 0.13 & 7.9 & - & 3 & (4),(7)\\
\hline
\noalign{\smallskip}
\end{tabular}
(1)=Abergel et al. 1996; (2)=AWM96; (3)=Bourke, Hyland \& Robinson 1995;
(4)=Benson \& Myers 1989; (5)=Caselli, Myers, Thaddeus 1995; (6)=Hilton \& Lahulla 1995; (7)=Keene 1981;
(8)=Loren 1989; (9)=Loren et al. 1990; (10)=Leung, Kutner \& Mead 1982; (11)=Parker 1988; (12)=Sandqvist \& Lindroos 1976, Sandqvist
1977; (13)=Schneider \& Elmegreen 1979; (14)=Tafalla et al. 1998; (15)=Ward-Thompson et al. 1994; (16)=Williams et al. 1999\\
$^{\dagger}$Oph D was also mapped in the LW3, LW6 and LW9 filters (see text and Fig.~\ref{absorption}cd). The zodiacal intensity is 45 MJy/sr in LW3.\\
$^{\ddagger}$Values of I$_{zodi}$ come from the zodiacal light model of Kelsall et al. (1998). Typical uncertainty is $\sim$ 20\%.\\
$^{*}$Cores belonging to Taurus molecular complex.\\
$^{**}$Cores belonging to $\rho$ Ophiuchi molecular complex.\\
\end{flushleft}
\end{minipage}
\label{sources}
\end{table*}

The
properties of the observed cloud cores are summarized in Table \ref{sources}. Column 1 lists the names of the sources as found in the dark
cloud catalogues, columns 2 and 3 list the J2000 coordinates of the centre of 
the ISOCAM field. The image centres are 
sometimes shifted from their original positions as given in the dark 
cloud catalogues (see above). Column 4 gives the
distances to the clouds. For the clouds not belonging to the $\rho$ Ophiuchi
(adopted distance d~=~160~pc\footnote{although recent studies suggest d~$\sim$~120$-$140~pc - see Knude
\& Hog (1998) and de Zeeuw et al. (1999).}) and
Taurus (d~=~140~pc) molecular complexes, we used the values from the study of Hilton \& Lahulla (1995) or the results
of Dame et al. (1987) on the locations of the major (giant) 
molecular clouds in the Galaxy to give an estimate of the
distances by relating the cores to known molecular complexes according to their galactic coordinates. Column 5 gives the ISOCAM filter(s) we used for the observations of each core. Column 7 lists the mean mid-IR intensity measured off the absorption features. Included in the values listed in column 7 is the zodiacal light emission (given in Column 8).
Column 9 lists the relative contrast between the intensity of the main absorption feature
and the mean mid-IR intensity (column 7), in percentage of the
mean mid-IR intensity
\footnote{Since the intensity in column 7 includes the contribution of the zodiacal light emission, the contrast in column 9 is not truly representative 
of the actual core absorption strength.}. 

Finally, column 10 gives the references of the dark cloud catalogues, from 
which our fields were selected, as well as the references of a 
few studies containing interesting information on the sources.
From Table \ref{sources}, it can be seen that the majority of the fields we observed 
contain absorption features. 
Figure~\ref{absorption} shows eleven cores with deep absorption  
(i.e., L1544, L1582A, Oph~D, R53, L1709A, R63, L1689B, L310, L328, 
L429, GF5 -- see Section \ref{isoimages})).

\subsection{ISOCAM observations}

\subsubsection{Observational parameters}

The dense cores were mapped between April 1997 and March 1998 using the LW2 (5--8.5 $\mu$m) and LW6 (7--8.5 $\mu$m) filters of ISOCAM
(Cesarsky et al. 1996), the spectro-imager aboard ISO (Kessler et al. 1996), centered at 6.75
$\mu$m and 7.75 $\mu$m respectively (except for Oph D, see below). 
These wavelengths approximately correspond to a minimum in the dust 
extinction curve 
(see the model of Draine \& Lee 1984 for diffuse clouds), providing a good opportunity to peer deep inside dark clouds.  
We used a pixel field of view of 6$\arcsec$, which also roughly corresponds to the
FWHM size of the ISOCAM point spread function (PSF) at $\sim$~7~$\mu$m. 
For a
``standard'' observation, an area of $10.7\arcmin \times 11\arcmin$ 
was covered for each core with
a raster map of $6 \times 4$ images and an overlap of 17 pixels in 
right ascension and of 6 pixels in
declination. At each raster position, about 18 single read-outs were 
taken with an integration time per read-out of 2.1s. 
About 50 stabilisation read-outs had to be added at the beginning of each observation to
allow the detectors to lose memory of the previous observation. For a few 
dense cores (L1672, SA187, and GF5), 
the field contained strong IRAS point sources and a shorter (0.28s)
integration time per read-out was adopted 
in order to avoid saturation on the camera detectors. 
In these cases, the number of read-outs was 44 at each raster position 
and the total area mapped was
$19.7\arcmin \times 18.8\arcmin$ for L1672 and SA187. 
In addition to the existing LW2 (5--8.5 $\mu$m) and LW3 (12--18 $\mu$m)
maps of Abergel et al. (1996, 1998) with 6$\arcsec$ pixels, 
we took two new images of Oph~D in the filters LW6 (7--8.5 $\mu$m) 
and LW9 (14--16 $\mu$m) with a pixel field of view of 3$\arcsec$. 
The Oph~D field was covered by a raster of $6 \times 3$ images with an 
overlap of 22 pixels in both directions. The raster was tilted by 20$\degr$ 
with respect to the Celestial North in order to follow the core's elongation (see Fig.~2d of MAN98 and Fig.~1 of Abergel et al. 1998). 
For these images of Oph~D, 15 exposures were taken at each raster position 
with an integration time of 10 sec per read-out. The total area covered 
was $2.6\arcmin \times 4.1\arcmin$.

\subsubsection{Data reduction}

The data were reduced using the CIA\footnote{`CIA' is a joint development by the ESA Astrophysics
Division and the ISOCAM Consortium. The ISOCAM Consortium is led by the ISOCAM PI, C. Cesarsky.} 
(CAM Interactive Analysis) 
software package. First, glitches due to cosmic rays were removed using
a spatial and temporal filtering algorithm (Starck et al. 1999), 
then a dark current image was subtracted from all the
frames. Because of the time lag in the response of the detectors, 
transient effects had to be
corrected for. For this purpose, we used the algorithm developed by 
Abergel et al. (1996).
Finally, the images were divided by a flatfield image to correct for the  non-uniform response of the detectors over the array. 
Most images were made up of over 500 frames that covered a large 
surface area. This enabled us to use a median image of all the frames 
as a flatfield image. 
Flatfielding is here a critical operation as we are looking for intensity
variations in dark objects and any flux mismatch between two contiguous raster images appears as a step in an intensity profile. Additional pixels had to be masked where the trails of glitches still affected the response of the 
detectors or induced a downward step in the flux level.

\subsubsection{Noise estimation}

In order to estimate the rms noise levels in our final images, 
we first derived the rms level of the temporal fluctuations for
each pixel of a given raster map, taking into account the overlap 
of the individual raster frames (each sky position in the map was observed 
on average in two independent individual frames). The resulting 
average rms noise per pixel depends on the level of the diffuse mid-IR  
emission in the raster, and thus varies from core to core. 
Typical values are $\sim$~0.05 MJy/sr for regions with low mid-IR
background (e.g. Taurus, Eastern part of the $\rho$~Oph complex) and 
$\sim$~0.1 MJy/sr for regions with high mid-IR background
(e.g. L328 and L310). 
This temporal rms noise however does not take flatfield noise 
(estimated to be $\sim$ 5\%) into account and could be
underestimating the total amount of noise. Another approach to estimating 
the noise in an image is
by calculating the rms noise level of the spatial fluctuations 
over regions showing very little structure. By doing so,
we a priori overestimate the noise level since we do not discriminate 
between true instrumental noise in the image and small-scale spatial 
structure in the (emitting and/or absorbing) source. 
The rms values listed in Table~\ref{sources} and 
the error bars used in the profiles of Sect.~\ref{coldensprof} below  
correspond to such spatial estimates and should thus be taken as
upper limits. Note that in the densest areas,
we can distinguish between small-scale structure and instrumental noise 
or mid-IR emission fluctuations by checking 
for the presence on small scales of cold,
IR-absorbing condensations in high-resolution millimeter 
continuum maps (see 2.3 below).


\subsection{Millimeter observations}

In order to check that the dark patches seen in the ISOCAM images 
(see Fig.~\ref{absorption}) were actually due to the presence of cold absorbing 
dense cores/condensations, 
and not to small-scale fluctuations of the mid-IR background and/or foreground, 
we carried out 1.3~mm continuum observations of a selected subset of 
our fields (see Table \ref{mm}). 
We used the IRAM 30~m telescope equipped with the 37-channel MPIfR 
bolometer array 
in March 1998 and March/April 1999 to map the 1.3~mm dust emission of 
those fields that presented clear and compact absorption-like features in
the ISOCAM images. 
On-the-fly 1.3~mm maps were taken in the dual-beam raster mode with a scanning 
velocity of 4$\arcsec$/sec, a spatial sampling of 2$\arcsec$ in azimuth 
and 4$\arcsec$ in elevation, and a chopping frequency of 2 Hz. 
The chop throw was set to 60$\arcsec$ for maps larger than 300$\arcsec$ 
and 45$\arcsec$ for smaller maps.
The (FWHM) beam size of the telescope was measured to be approximately
11$\arcsec$-12$\arcsec$ on maps of Uranus. 
The zenith atmospheric optical depth, monitored by `skydips' 
every 1-2 hours, was between $\sim$ 0.15 and $\sim$ 0.5.
The final co-added maps are about 7$\arcmin \times 7\arcmin$ in size and 
consist of $\sim$~5--6 individual coverages taken at different
hour angles 
to improve the reconstruction technique. 
The data were reduced with the IRAM software for
bolometer arrays ("NIC"; cf. Brogui\`ere, Neri \& Sievers 1995), which uses the
EKH restoration algorithm (Emerson, Klein, \& Haslam 1979). The typical 
rms noise was $\sim$~8~mJy/11$\arcsec$beam.


The same dense cores listed in Table \ref{mm} were also observed in the
C$^{18}$O(1-0) molecular line at 
the IRAM 30~m telescope, in order to obtain estimates 
of the column density in the outer parts of the cores.
We used the position switching mode along
cuts of about 10$\arcmin$ length sampled every 20$\arcsec$ through each of the cores, and oriented along and perpendicular to the axis of the core. 
The OFF position, which was checked to be free of
emission, was typically taken between 15$\arcmin$ and 30$\arcmin$ away from the nominal source position. 
Our goal was to get good signal-to-noise ratio in the outer parts of 
the core in order to estimate the H$_{2}$ column density, $N_{H_{2}}$, 
as far from the core centre as possible (typically $\sim$~200$\arcsec$). 
The total (ON+OFF) integration time was 2 minutes on average and up to 10
or 25 minutes in the outer parts of the cores in order to reach an rms sensitivity better than $\sim$~0.1~K~km~s$^{-1}$. The spectral resolution was 
20~kHz and the observing frequency 109~GHz.  
Additional N$_{2}$H$^{+}$(1-0), H$_{2}$CO(2-1), and DCO$^{+}$(2-1) spectra
were taken at a few positions for selected cores. 

During these continuum and line runs at the 30~m telescope, 
the pointing, as well as 
the focus, was checked every $\sim$ 1 hour and found to be better 
than $\sim 5\arcsec$ in both azimuth and elevation. 

  
\section{Results and analysis}

\label{images}

\begin{figure*}[htpb]
\psfig{file=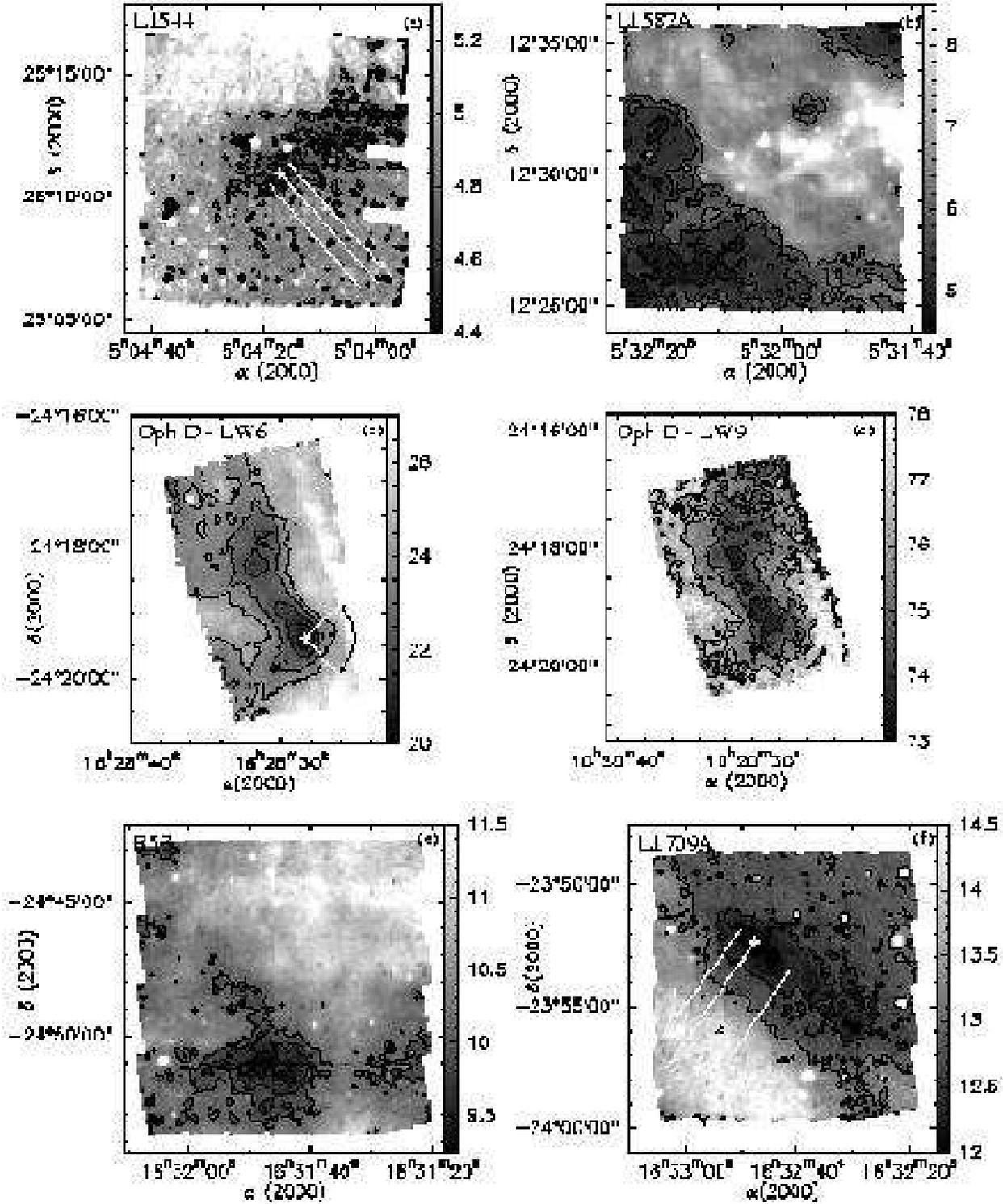,width=17.5cm}
\caption{ISOCAM images of the sample of 11 dense cores presenting strong absorption features (see text). 
The cores appear as dark patches against a brighter background
emission. The ISOCAM psf is $\sim$~6$\arcsec$ (FWHM) at $\sim$~7~$\mu$m; the pixel size is 
also 6$\arcsec$ (3$\arcsec$ in the case of Oph~D). 
White crosses and triangles show the core centers and positions where 
the C$^{18}$O column density was measured (section \ref{outercoldens}), 
respectively. The dotted lines/sectors illustrate the way averages were performed (section \ref{coldensprof}). 
The greyscale is indicated in MJy/sr at the right of each image. 
Contour levels are: a) for L1544, 4.55, 4.65, 4.75 MJy/sr; b) for L1582A, 
5, 5.5, 6 MJy/sr; c) for Oph D LW6 (7.75 $\mu$m), 21, 22, 23, 24 MJy/sr;  d) for Oph D LW9 (15 $\mu$m), 74, 75, 76 MJy/sr; e) for
R53, 9.5, 9.75, 10 MJy/sr; f) for L1709A, 24, 25, 26 MJy/sr.}
\label{absorption}
\end{figure*}
\setcounter{figure}{0}
\begin{figure*}[htbp]
\psfig{file=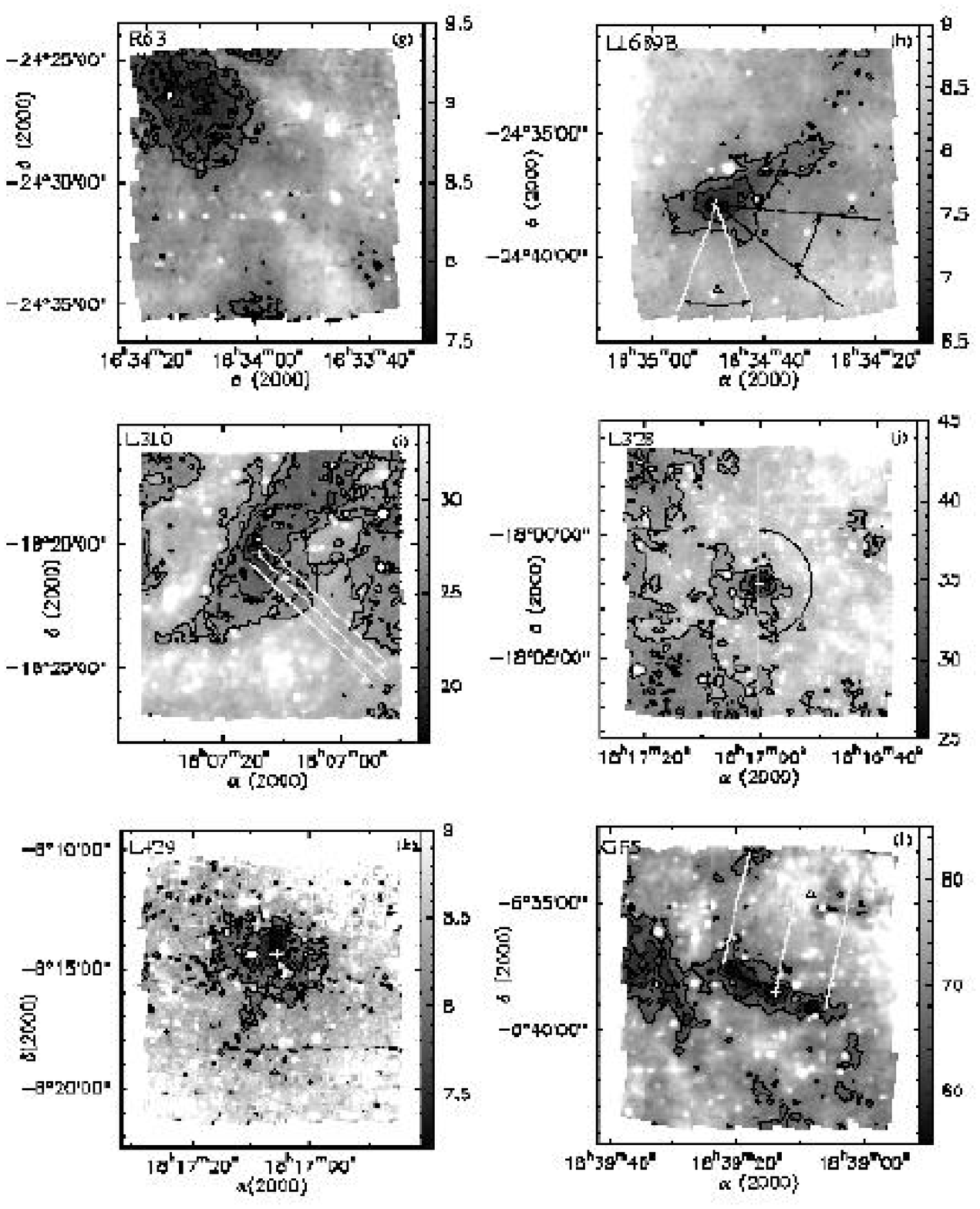,width=18cm}
\caption{(cont'd) Contour levels:  
g) for R63, 7.75, 8, 8.25 MJy/sr; h) for L1689B, 7, 7.5, 8
MJy/sr;  i) for L310, 20, 22.5, 25, 27.5 MJy/sr; j) for L328, 30, 34, 38 MJy/sr; k) for L429,
7.3, 7.7, 8.1, 8.5 MJy/sr; l) for GF5, 57, 62, 67 MJy/sr.}
\end{figure*}
\begin{figure*}[htbp]
\psfig{file=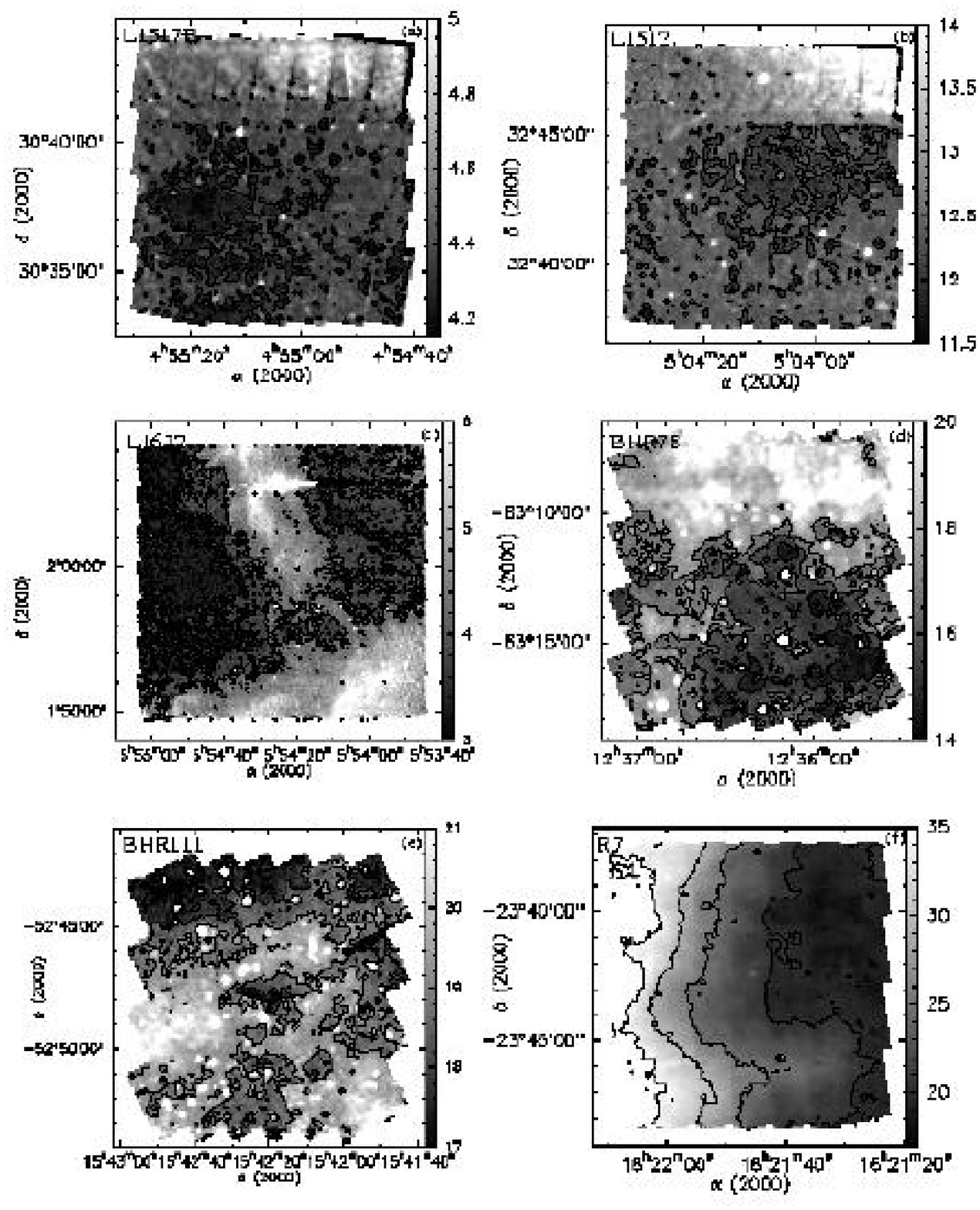,width=18cm}
\caption{ISOCAM images of the remaining cores. The fainter absorption features 
led us to enhance the contrast of the images, which resulted in poorer image 
appearance (the flatfield noise is clearly visible in some cases). Contour levels are: 
a) for L1517B, 4.3 MJy/sr; b) for L1512, 12.2 MJy/sr; c) for L1672, 3.5, 4 MJy/sr; d) for 
BHR78, 15, 16, 17; e) for BHR111, 17, 18, 19 MJy/sr; f) for R7, 20, 25, 30, 35 MJy/sr.}
\label{weak}
\end{figure*}
\setcounter{figure}{1}
\begin{figure*}[htbp]
\psfig{file=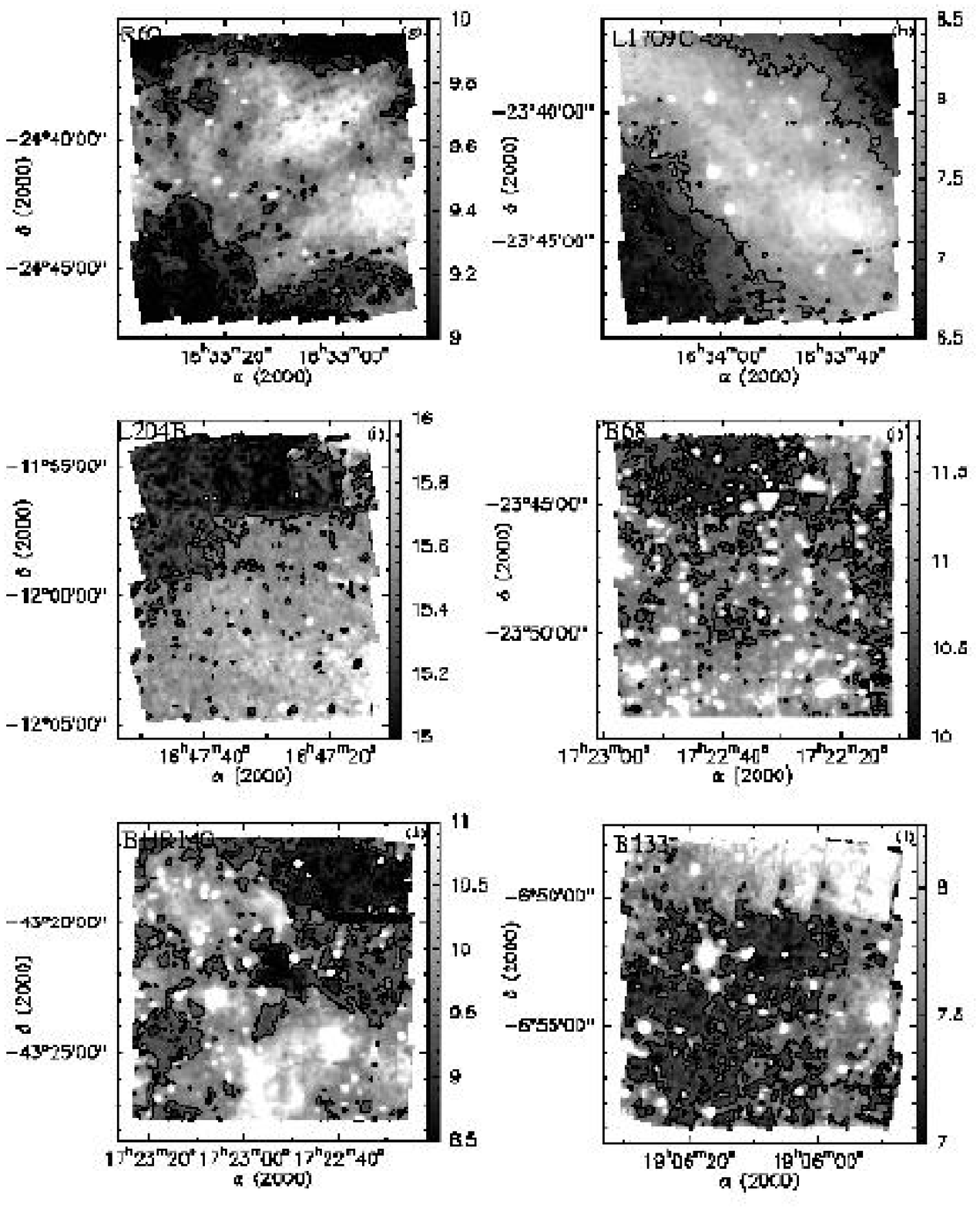,width=18cm}
\caption{(cont'd) Contour levels: g) for R60, 9.1, 9.3 MJy/sr; h) for L1709C, 6.9,
7.4 MJy/sr; i) for L204B, 12.2 MJy/sr; j) for B68, 5.4, 5, 6 MJy/sr; k) for BHR140, 9,
9.5 MJy/sr; l) for B133, 7.4 MJy/sr.}
\end{figure*}

\subsection{ISOCAM images}

\label{isoimages}

Out of the 24 dark cloud fields in our programme, 
23 show  absorption-like features in the mid-IR. 
\begin{figure*}[htbp]
\psfig{file=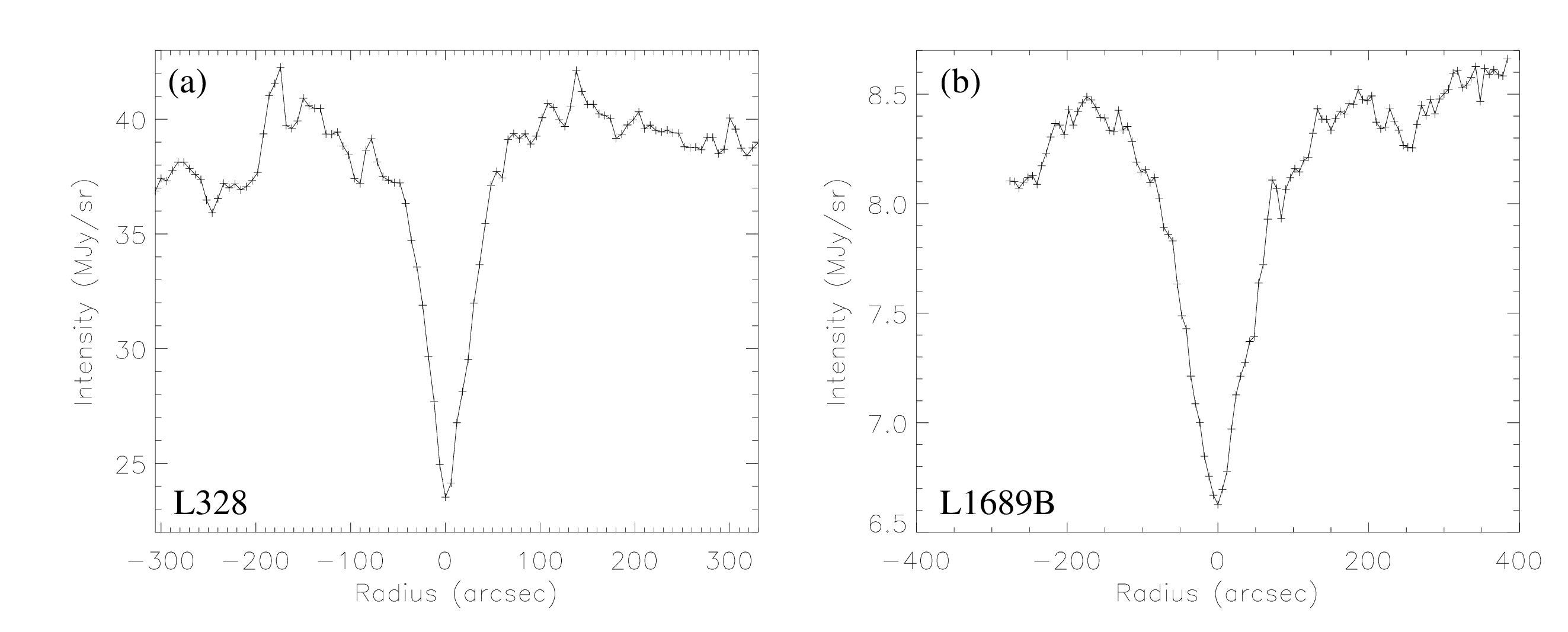,width=18cm}
\caption{(a) North-South intensity cut through the $\lambda = 7.75$~$\mu$m 
ISOCAM image of L328 (see Fig. \ref{absorption}), 
at $\alpha \sim 18^h 17^m 00\fs$ 
(b) North-South intensity cut through the $\lambda = 6.75$~$\mu$m ISOCAM image of L1689B at $\alpha \sim 16^h 34^m 50\fs$
Both cloud cores clearly show up in absorption against the diffuse 
mid-infrared background.}
\label{cut}
\end{figure*}

The fields presenting such features can be divided into two broad categories, 
those that show deep ($\simgt$ 10\% - cf. column 9 of Table~\ref{sources}), compact absorption features likely associated with 
dense cores, and those presenting weaker ($\simlt$ 10\%) and more diffuse absorption whose nature 
is also more uncertain.  
Figure \ref{absorption} presents 
the ISOCAM images of 11 fields with strong absorption features, to which we
added the widely studied core L1544 (e.g. Ward-Thompson et al. 1994, Tafalla et al. 1998, Williams et al.
1999, Ohashi et al. 1999), and from which we withdrew L1709C (the absorption features lying in the corners of the image, 
making the analysis difficult). 
The remaining fields are presented in Fig. \ref{weak}. 
Figure \ref{cut} shows examples of cuts through two of the deep
absorption features (for L328 and L1689B). 
The morphologies encountered in the fields with compact 
absorption vary from spherical cores like L328 to more elongated, 
filamentary structures (e.g. L1689B, GF5).

\subsection {Comparison with the millimeter continuum maps}

All of the cores observed at 1.3~mm (Table~\ref{mm})
present extended, weak dust continuum emission which we were able to map. For all cores
, the peak of the mid-IR absorption 
does correspond with a peak in the millimeter emission. 
This is illustrated in Fig.~\ref{isomm} which shows the mid-IR ISOCAM 
absorption contours superimposed on the 1.3~mm continuum maps of the eight 
cores positively detected at 1.3~mm (the maps of Oph D and L1544 are also shown in MAN98 and WMA99, respectively).

\begin{figure*}[htbp]
\psfig{file=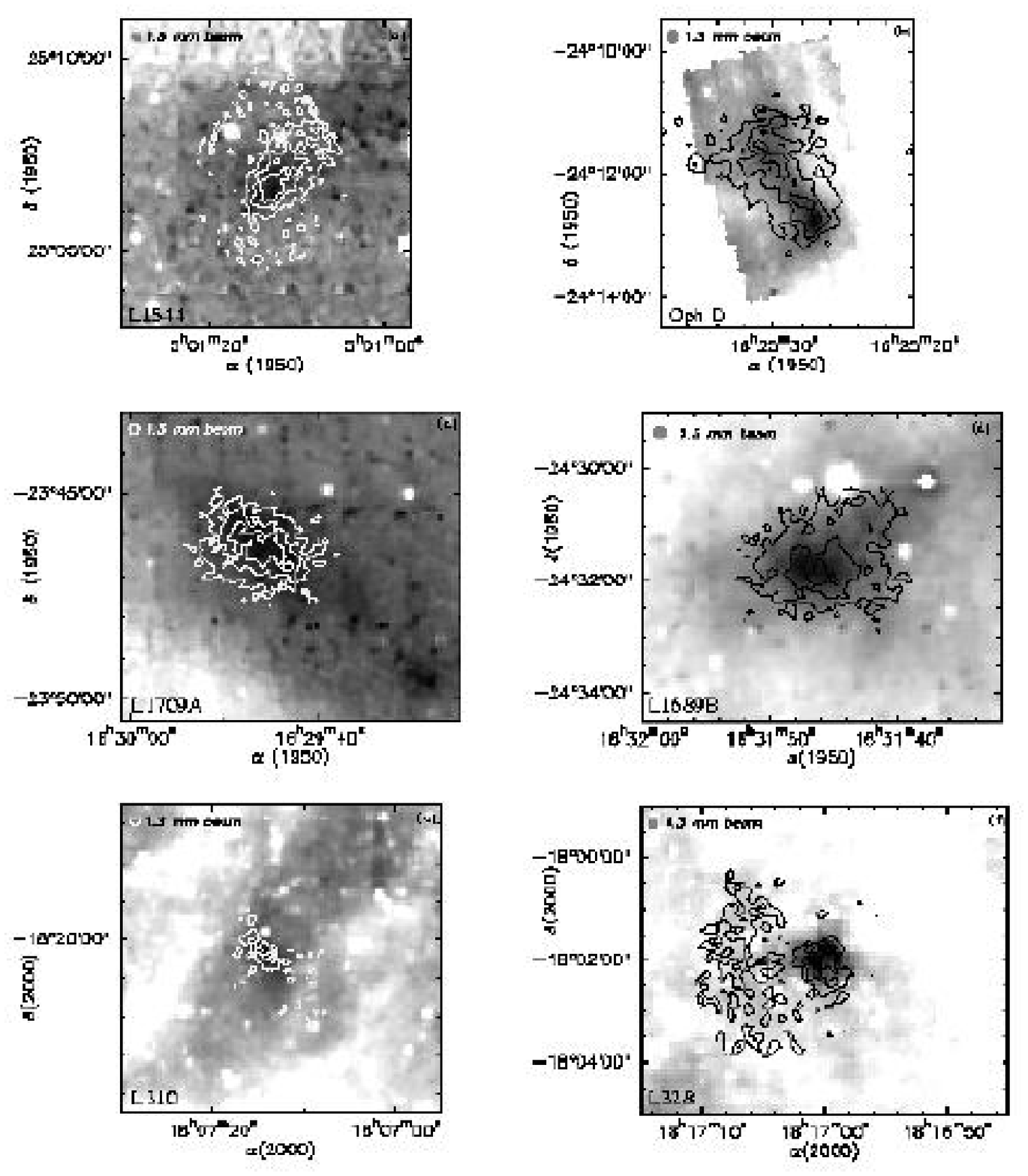,height=21.0cm,angle=0}
\caption{IRAM 1.3~mm continuum emission maps smoothed to 13$\arcsec$ FWHM resolution (contours) superimposed on ISOCAM
7~$\mu$m absorption image (greyscale). Contour levels are: 
a) 20, 40, 60 mJy/beam for L1544; b) 30, 50, 70 mJy/beam for Oph D; 
c) 20, 40, 60 mJy/beam for L1709A; d) 10, 30, 50  mJy/beam for L1689B; 
e) 20, 40, 60  mJy/beam for L310; f) 20, 40 mJy/beam for L328.}
\label{isomm}
\end{figure*}

\setcounter{figure}{3}
\begin{figure*}
\psfig{file=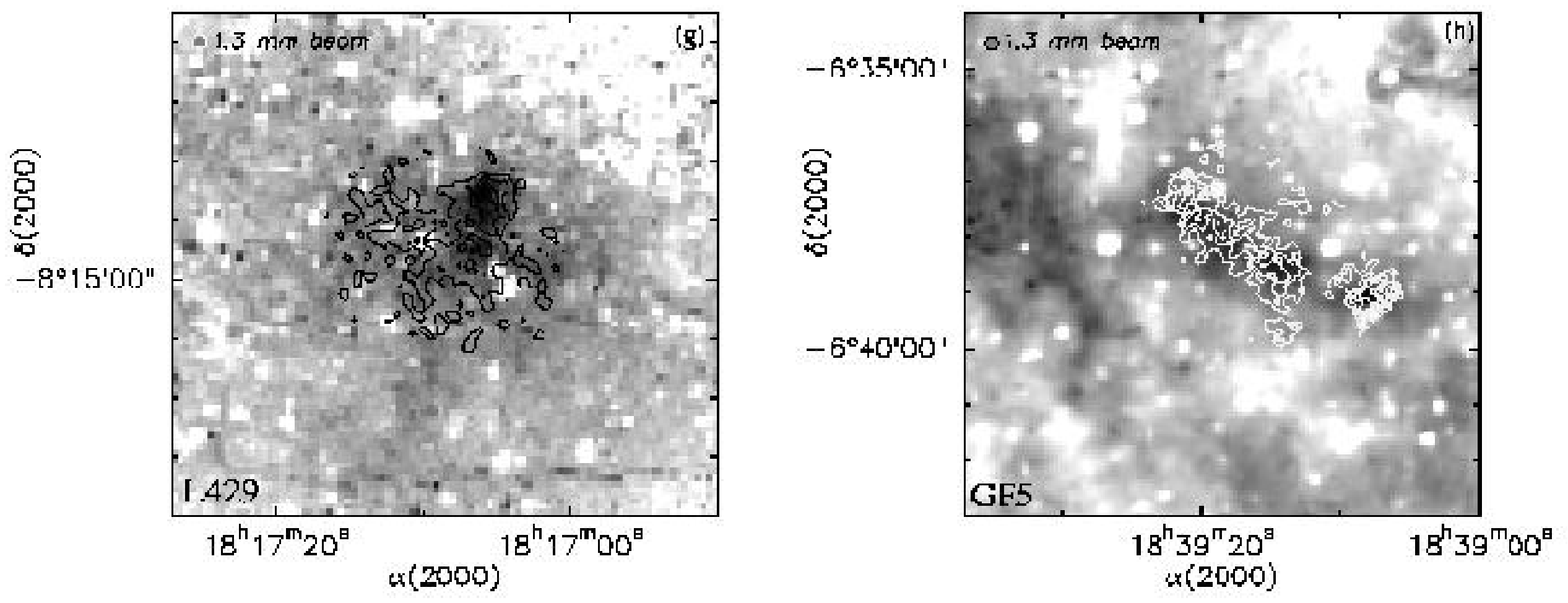,height=7.5cm,angle=270}
\caption{(cont'd) Contour levels are: g) 25, 50, 75, 100 mJy/beam for L429;
h) 10, 20, 30, 40 mJy/beam for GF5.}
\end{figure*}

The good agreement between the respective 
images of these cores in both types of data confirms that the
features we see with ISOCAM are due to absorption by the (rather large)
column density of cold dust traced at 1.3~mm, and not by fluctuations in 
the mid-IR background.
Furthermore, the similar morphologies seen in both wavebands suggest that 
the mid-IR absorption images trace the same dust as the millimeter emission maps, 
which justifies a detailed comparison of the column density profiles derived 
from both tracers.

\begin{table*}[ht]
\caption[]{Core parameters deduced from millimeter observations.}
\begin{minipage}[b]{14cm}
\begin{tabular}{lcccccccc} \hline
Name & $R_{flat}$ & $N_{H_{2}}^{flat}$ & $R_{out}$ & $N_{H_{2}}^{out}$ & $I_{zodi}^{MIR}$ & $I_{fore}^{MIR}$ & $I_{back}^{MIR}$  \\
 & (AU) & cm$^{-2}$ & (pc) & cm$^{-2}$ & (MJy/sr) & (MJy/sr) & (MJy/sr) \\
     &           (1) & (2) & (3) & (4) & (5) & (6) & (7)\\
\hline
\hline
L1544 & 1900 & 4.5 $\times 10^{22}$$^{\dagger}$ & 0.08 & 2$\times 10^{21}$ & 4.2 & 4.2$^{+0.2}_{-0.7}$ & 0.7$^{+0.7}_{-0.2}$  \\
Oph D (LW2) & 3400 & 5$\times 10^{22}$ & 0.05 & 3.2$\times 10^{21}$ & 5.7 & 5.7$^{+1.5}_{-1.1}$ & 9.1$^{+1.0}_{-1.3}$ \\
Oph D (LW3) & ``        &       ``      & ``    & ``            &       45 & 42.1$^{+2.6}_{-1.0}$ & 9.5$^{+2.0}_{-2.4}$ \\
L1709A & 6400 & 4.5$\times 10^{22}$ & 0.18 & 3.1$\times 10^{21}$ & 10.5 & 11.3$^{+0.6}_{-1.3}$ & 1.7$^{+1.3}_{-0.6}$ \\
L1689B s  & 4000 & 4$\times 10^{22}$ &0.16 & 2$\times 10^{21}$ & 4.3 & 4.5$^{+1.2}_{-1.0}$ & 3.8$^{+1.0}_{-1.2}$ \\
L1689B w  & 4000 & 4$\times 10^{22}$ & 0.29 & 4.8$\times 10^{21}$ & 4.3 & 4.3$^{+1.0}_{-0.4}$ & 4.3$^{+0.3}_{-0.7}$ \\
L310 & 8000 & 1.7$\times 10^{22}$ & 0.11 & 1$\times 10^{21}$ & 7.1 & 7.1$^{+0.3}_{-1.2}$ & 19.2$^{+1.1}_{-0.2}$ \\
L328 & 5000 & 3$\times 10^{22}$ & 0.14 & 2.4$\times 10^{21}$ & 7.4 & 7.4$^{+6}_{-1.5}$ & 34.5$^{+0.9}_{-5.3}$ \\
L429 &  2400 & 5$\times 10^{22}$ & 0.26 & 8$\times 10^{21}$ & 6.8 & 6.8$^{+0.2}_{-1.4}$ & 2.3$^{+1.3}_{-0.1}$ \\
GF5 & 5000 & 2$\times 10^{22}$ & 0.26 & 2$\times 10^{21}$ & 7.2 & 9.1$^{+17.6}_{-2.5}$ & 70.8$^{+1.5}_{-16.6}$ \\
\hline
\end{tabular}
Notes to Table \ref{mm}:

(1) $R_{flat}$ is the radius of the flat inner part of the core column density profile (see WMA99 and Sect. \ref{coldensprof}). For L1544, Oph D and L1689B, $R_{flat}$ was taken from the millimeter continuum measurements of WMA99 and corresponds to the geometrical average of the major and minor axes of the ellipse. For L1709A, L310, L328, L429 and GF5, $R_{flat}$ is the radius of the flat inner part of the ISOCAM profile.\\
(2) $N_{H_{2}}^{flat}$ is the column density averaged over the flat inner region of the core of radius $R_{flat}$ as deduced
from our millimeter continuum measurements (see section \ref{innercoldens}).\\
(3) $R_{out}$ is the radius at which the C$^{18}$O column density was measured in the outer parts of the core (see section \ref{outercoldens}).\\
(4) $N_{H_{2}}^{out}$ is the column density inferred from C$^{18}$O(1-0) observations at $R_{out}$ (see section \ref{outercoldens}).\\
(5) $I^{MIR}_{zodi}$ is the zodiacal light emission in the mid-IR (see col. [8] of Table \ref{sources}).\\
(6), (7) $I_{back}^{MIR}$ \& $I_{fore}^{MIR}$ are the mid-IR background and foreground intensities as deduced from Eq. (\ref{itau}) and the values of the central and outer column
densities, using the additional constraint that $I_{fore}\geq I_{zodi}$ (see section \ref{intensities}).\\
$^{\dagger}$The values of $N_{H_{2}}^{flat}$ and $<N_{H_{2}}>_{13\arcsec}$ given in Table 2 of WMA99 are too large by $\sim$15 \% and $\sim$~70~\%, respectively.

\end{minipage}
\label{mm}
\end{table*}

\subsection {Modelling of the mid-infrared absorption}

\label{modelling}

In order to derive information about the column density structure from the 
mid-IR intensity, we have used a simple geometrical model: each 
dense core is embedded within a lower-density parent molecular cloud. 
On the line of sight to the core, we 
measure i) a background intensity arising from the rear side of 
the parent cloud that is attenuated by the absorption from the core,  
and ii) a foreground intensity arising from the front side of the parent 
cloud. We can thus express the mid-IR intensity along the line of sight as:
\begin{eqnarray}
        I(\bar{r}) & = & [<I_{back}>  + \delta I_{back}(\bar{r})] \cdot 
          e^{-\tau_{\lambda}(\bar{r})}\nonumber\\ 
&  & + [<I_{fore}> + \delta I_{fore}(\bar{r})],
\label{itau}
\end{eqnarray}
where $I_{back}$ and $I_{fore}$ 
are the background and foreground intensities respectively, 
and $\tau_{\lambda}(\bar{r})$ is the dust opacity 
at a projected radius $\bar{r}$ from core center and at observed wavelength 
$\lambda$. The foreground emission, $I_{fore}(\bar{r})$, 
includes the contribution from the zodiacal light 
emission, $I_{zodi}$.


The mid-infrared background emission, $I_{back}(\bar{r})$, is thought to arise 
from very small grains presenting aromatic emission features (L\'eger \& Puget 1984, Allamandola, Tielens \& Barker 1989, Boulanger et al. 1996). 
These grains/molecules are excited by the interstellar far-UV (FUV) 
radiation field and re-emit this energy as a group of spectral lines/bands 
in the mid-infrared, the Unidentified Infrared Bands (UIBs). 
The ISOCAM LW2=[5.0, 8.5 $\mu$m] and LW6=[7.0, 8.5 $\mu$m] filters that 
we used for our observations include the 6.2 $\mu$m 
and 7.7 $\mu$m emission features of these UIBs.

According to recent studies, the small particles responsible for the UIBs are distributed 
throughout molecular clouds
but because of high interstellar extinction in the FUV band, 
they are ususally only excited in the outer regions 
where $A_{v}\simlt 1$ (Hollenbach \& Tielens 1997, Bernard et al. 1993).
For this reason, the interiors of (quiescent) clouds should be relatively 
free of UIB emission,  
and only their outer envelopes should significantly contribute to the mid-IR 
background and foreground (e.g. Bernard et al. 1993, Boulanger et al. 1998). 

As a first step, we neglected the spatial fluctuations of the background 
and foreground, $\delta I_{back}(\bar{r})$ and $\delta I_{fore}(\bar{r})$, and thus 
assumed $I_{back}$ and $I_{fore}$ to be equal to their mean values 
across the field, $<I_{back}>$ and $<I_{fore}>$ 
(see \ref{nonuni} below for a critical discussion of this assumption). 
 
For a given gas-to-dust ratio, the H$_{2}$ column density $N_{H_{2}}$ is 
related to the dust opacity by 
        $\tau_{\lambda}(\bar{r})=\sigma_{\lambda} \cdot N_{H_{2}}(\bar{r})$, 
where $\sigma_{\lambda}$ is the dust extinction cross section\footnote{At the observed mid-IR wavelengths, the scattering component of $\sigma_{\lambda}$ should be more than ten times smaller than the absorption component for a dust composition following Preibisch et al. (1993) 
and can thus be neglected (e.g. Bohren \& Huffman 1983).} 
at the wavelength of observation $\lambda$. 
We assume  $\sigma_{\lambda}$ to follow the Draine \& Lee (1984) dust model 
in the mid-IR as a first approximation, but the value is rather 
uncertain (see 
section \ref{uncert}). We adopted $\sigma_{6.75\mu m}=1.2\times 10^{-23}$ cm$^{2}$ for the LW2 filter, $\sigma_{7.75\mu m}=1.35\times 10^{-23}$ cm$^{2}$ for the LW6 filter, and $\sigma_{15\mu m}=1.6\times 10^{-23}$ cm$^{2}$ for the LW9 filter, i.e., we calculated 
$\sigma_{\lambda}$ at the central wavelength of each filter. A more accurate value could be obtained by integrating the extinction curve over the 
response of the 
ISOCAM filters, but as we will see in the following, the exact value of
$\sigma_{\lambda}$ has only little influence on the derived 
column density profiles. 

The quantities $I_{back}$ and $I_{fore}$ are unknown, but they can be constrained using independent measurements of the H$_{2}$ column density at 
two different radii. These constraints enable the ISOCAM absorption profile 
to be converted into a column density profile. 



\subsubsection {Central column density from millimeter continuum data}

\label{innercoldens}

We used our 1.3 mm continuum maps to estimate the column density averaged 
over the flat central region of each core.
As millimeter dust continuum emission is optically thin, 
the column density averaged over the flat inner part (of radius $R_{flat}$) of the core (see WMA99)
may be derived from the  1.3~mm flux density integrated over the same area, $S_{1.3mm}^{flat}$ (cf. AWM96, MAN98):
\begin{equation}
N^{flat}_{1.3mm}=S^{flat}_{1.3mm}/[\Omega^{flat} \kappa_{1.3mm} \mu m_{H} B_{1.3mm}(T)],
\label{nflat}
\end{equation}
where $\Omega^{flat}$ is the solid angle of the flat inner region, 
$\kappa_{1.3mm}$ is the dust opacity per unit mass column density at $\lambda$~=~1.3~mm, $\mu=2.33$ is the mean molecular weight, and $B_{1.3}(T)$
is the Planck function at $\lambda$~=~1.3~mm, for a dust temperature $T$. 
We assumed a single, representative dust temperature $T_d = 12.5$~K for  
all the cores. Based on recent ISOPHOT measurements of L1544, Oph-D, L1689B, 
and other similar starless cores (e.g. Ward-Thompson \& Andr\'e 1999, Lehtinen 
et al. 1998), we believe that this value of $T_d$ is likely to be within 
$\pm 3$~K of the true dust temperature in most cases. 
Considering the additional uncertainty on the millimeter 
dust mass opacity in dense cores, 
which we took equal to $\kappa_{1.3mm} = 0.005$~cm$^{2}$g$^{-1}$ 
(see, e.g., Preibisch et al. 1993, AWM96, and Kramer et al. 1999), 
the estimates of $N_{H_{2}}^{flat}$ listed in Table~2 are 
uncertain by a factor of $\simgt 2$ on either side of their nominal values. 

\subsubsection {Outer column density from millimeter line data}

\label{outercoldens}
 
The second value of the column density we can use to calibrate the ISOCAM 
column density profile is given by our C$^{18}$O(1-0) 
line observations of the outer parts of the cores. 
Assuming optically thin emission and Local Thermodynamical Equilibrium (LTE), 
we can determine the value of the C$^{18}$O column density from 
the following equation:


\begin{equation}
 N_{C^{18}O} (cm^{-2}) = 4.76\times 10^{13} 
\frac{T_{k}+0.88}{\exp(-\frac{5.27}{T_{k}})}\int_{-\infty}^{+\infty}T_{A}^*(v)dv,
\end{equation}
where $T_{A}^*$ is the measured antenna temperature 
of the C$^{18}$O(1-0) line, $T_{k}$ is 
the gas kinetic temperature in the cloud, and $dv$ is in km~s$^{-1}$ (e.g. Rohlfs \& Wilson
1996).
We used Eq.~(2) with $T_{k} = 10$~K to estimate the C$^{18}$O column density 
in the outer parts of the core. This method would not be reliable towards the center 
of the core where C$^{18}$O(1-0) may be optically thick. 
Moreover at the centre, the possibility of depletion of
gas molecules onto grains makes the estimation of the column density unreliable
(e.g. Kramer et al. 1999).
We adopted the following relation between H$_{2}$ and C$^{18}$O 
column densities (Frerking, Langer \& Wilson 1982): 
$N_{H_{2}}=5.53 \times 10^{6} N_{C^{18}O} + 1.22 \times 10^{21}$~cm$^{-2}$.
The values of the column density $N_{H_{2}}^{out}$ derived in this way, as well as the distance $R_{out}$
from the core centre at which it was evaluated, are given in Table~\ref{mm}. 
The exact position where $N_{H_{2}}^{out}$ 
was measured is indicated by a triangle-symbol on the images of 
Fig. \ref{absorption}.
The uncertainty on the H$_{2}$ column density derived from molecular line 
data has two major
origins: first, the excitation state of the C$^{18}$O(1-0) molecules is not accurately known and the LTE hypothesis is only an approximation 
(cf. White et al. 1995), and second, the C$^{18}$O abundance ratio with 
respect to H$_{2}$ is also uncertain in our cores. 
Based on comparisons with more sophisticated calculations 
performed with a Monte-Carlo radiative transfer code (Blinder 1997) 
for a realistic pre-stellar core model, we judge 
that our LTE estimate of 
the column density in the outer parts of each core, $N_{H_{2}}^{out}$,
is good to better than a factor of $\sim 2$.

\subsubsection {Estimates of the background and foreground intensities}

\label{intensities}

Given the values of $N_{H_{2}}^{flat}$ and $N_{H_{2}}^{out}$ derived 
from millimeter observations, it is easy to estimate $I_{back}$ and 
$I_{fore}$ using Eq~(\ref{itau}) for $\bar{r} = R_{flat}$ and $\bar{r} = R_{out}$,
and then deduce the column density
from the mid-IR intensity at any position in the ISOCAM image. 
The values of $I_{back}$ and $I_{fore}$ that we calculated in this way, 
using the independent constraints that $I_{back} \geq 0$ and 
$I_{fore} \geq I_{zodi}$, are listed in Table \ref{mm}. 
In addition to `best' values, we give permitted ranges for $I_{back}$ 
and $I_{fore}$ that were determined using the maximum and
minimum values of $N_{H_{2}}^{out}$ and $N_{H_{2}}^{flat}$ (given the 
factor $\sim 2$ uncertainty on both these quantities). 

Another approach consists in estimating the values of $I_{back}$ and $I_{fore}$ directly from the ISOCAM images. The sum $I_{back}+I_{fore}$ is constrained 
by the mean mid-IR intensity $<I_{MIR}>$ measured in the map outside the 
dense core (cf. column~7 of Table~\ref{sources}; this also corresponds 
to the ``baseline level'' measured on intensity cuts such as those 
of Fig.~\ref{cut}). Furthermore, one can set straightforward 
lower and upper limits on the foreground intensity since 
$I_{fore}$ has to be less than the value of the minimum intensity in the 
image, $I_{min}$, and greater than the zodiacal emission, $I_{zodi}$. 
Hence, $I_{zodi}<I_{fore}<I_{min}$ and, since $I_{back}+I_{fore}=$~~$<I_{MIR}>$, 
~~$<I_{MIR}>-I_{min}$~~$<I_{back}<$ $<I_{MIR}>-I_{zodi}$.

This second method assumes that $<I_{MIR}>$ can be measured 
outside the dense core, which is not necessarily the case if, e.g., the core 
extends beyond the limits of the ISOCAM map. 
Moreover, according to Eq.~(1), 
the minimum intensity in the image is equal 
to the intensity at the maximum of the absorption
($I_{min}= <I_{back}> \cdot e^{-\tau_{max}} + <I_{fore}>$), 
which in practice can greatly overestimate the value of $<I_{fore}>$
since one typically has $\tau_{max} \simlt 1 $ (see below).
For these reasons, we preferred to adopt
the first approach based
on millimeter measurements of the central and outer column densities
(see above).

In the case of the Oph~D dense core, however, it is possible to
determine accurate values of $I_{back}$ and $I_{fore}$ at more than 
one wavelength using both methods,
since large-scale ($\sim 0.8\degr\times 0.8\degr$) ISOCAM images of the
$\rho$ Oph main cloud are available (cf. Abergel et al. 1996).
This provides an important consistency check.
At 6.75~$\mu$m, we obtain
$I_{fore}^{6.75} = 5.7^{+1.5}_{-1.1}$~MJy/sr and 
$I_{back}^{6.75} = 9.1^{+1.0}_{-1.3}$~MJy/sr using millimeter observations
(see Table \ref{mm}), while we find $4.6$~MJy/sr $ <I_{fore}^{6.75}<8.5$~MJy/sr and
$6.3$~MJy/sr $<I_{back}^{6.75}<10.2$~MJy/sr using the second method. The agreement is thus very good.
The background and foreground intensities 
can also be estimated at $\sim 15$~$\mu$m (cf. Table \ref{mm}).
As can be seen in Fig.~\ref{profiles}c) and \ref{profiles}d) below, they yield a column density profile
for Oph~D which is virtually identical to that derived at 6.75~$\mu$m.
Furthermore, our estimated values ($I_{fore}^{15} = 42.1^{+2.6}_{-1.0}$~MJy/sr and  
$I_{back}^{15} = 9.5^{+2.0}_{-2.4}$~MJy/sr) are consistent with the idea 
that the mid-IR background is primarily due to the emission of 
UIB carriers while the foreground is dominated by the zodiacal light.
Indeed, the emission spectrum of aromatic particles is such that 
the ratio of the LW2-filter to LW3-filter intensities is $\sim 1$ 
(Boulanger et al. 1996), which is precisely what we find for the background 
emission in Oph D. By contrast, the 7--15~$\mu$m spectrum of 
the foreground emission is rising and compatible with that 
of the zodiacal light ($I_{zodi}\sim 5.7$~MJy/sr at $\sim 7$~$\mu$m and $I_{zodi}\sim 45$~MJy/sr 
at $\sim 15$~$\mu$m).

The results of these inter-comparisons on Oph~D confirm the validity of 
the millimeter method we used to determine $I_{back}$ and $I_{fore}$ 
for the other cores.

We note that for many sources $I_{fore} \sim I_{zodi}$ in Table~\ref{mm}, 
suggesting that the zodiacal emission is often 
the main contributor to the foreground. This situation is expected if most 
of the target clouds are illuminated from behind, which is indeed believed 
to be the case for the $\rho$~Oph cloud, located in front of the Sco~OB2 
association (e.g. de Geus et al. 1989, de Geus 1992). As most of our 
fields were chosen to be nearby and devoid of strong (e.g. foreground) 
infrared sources, the fact that the front sides of the clouds appear to 
contribute negligible foreground emission is likely the result of a selection
effect.


\subsection{Radial column density profiles}

\label{coldensprof}

\subsubsection{Derivation of the profiles}

In order to characterize the radial structure of each core and reduce 
the influence of spatial fluctuations in the background and foreground,
we first derived mean radial intensity profiles by 
averaging the mid-IR emission in the ISOCAM images according to the 
apparent morphology of the core. 
For L1689B, we averaged the intensity over elliptical annuli of increasing 
radii separately for a 40\degr sector in the South and a 40\degr sector in 
the Western side of the core. 
For L328, OphD and L429, the intensity was averaged over circular annuli
of increasing radius. Only the South-Western sector of the profile was considered for Oph D and only the
Western half for L328. For very elongated, 
filamentary structures (L310, L1709A, L1544, GF5), we averaged cuts perpendicular to the main axis of the 
cores, but only kept the Southern half of the absorption intensity profile (except GF5 for which the North side was kept)\footnote{
For Oph D, the Eastern profile does not stretch far enough to enable an analysis at large radii ($>$~200$\arcsec$).
For L328, 
the Eastern profile is disturbed by the presence of
additional absorption features. For L310, L1709A, L1544 and GF5, the profile we kept was
the longest one not affected by the presence of supplementary absorption features
(or by flatfield noise as in L1544). }. 
The sectors over which the profiles were averaged (for the elliptically/spherically symmetric cores L1689B, L328 and L429) are shown as dotted lines on Fig. \ref{absorption} as are the ranges of cuts which were averaged for the filamentary cores (L1544, L1709A, L310, GF5).

The radial column density profiles $N_{H_{2}}^0(\bar{r})$
that were derived in this way using 
our `best' values for $I_{back}$ and $I_{fore}$ (see above) 
are displayed in Fig.~\ref{profiles} as crosses. 
The vertical cross sizes represent statistical error bars on the mean radial profile, 
estimated from the dispersion of the values averaged at each radius, divided by the square root of the number of values. 
In addition to these random errors at each radius, there is also a 
global uncertainty on the column density profiles resulting from our 
uncertain knowledge of $I_{back}$ and $I_{fore}$.
The dashed curves lying above and below the `best' profiles limit the range 
of column density profiles that are consistent with the mid-IR absorption 
data, given the permitted range of values for $I_{back}$ and $I_{fore}$ in 
each case (see Table~2). The upper dashed curve corresponds to the profile 
for the smallest value of $I_{back}$ and the greatest value of $I_{fore}$, 
while the lower dashed curve corresponds to 
the greatest value of $I_{back}$ and the smallest value of $I_{fore}$.
It is important to stress that there is a simple, global relation 
between each permitted profile, $N_{H_{2}}(\bar{r})$, 
and the `best' profile, $N_{H_{2}}^0(\bar{r})$, of the type:
\begin{displaymath}
 N_{H_{2}}(\bar{r})= 
  \frac{1}{\sigma_{\lambda}}\cdot\ln\biggl(\frac{1}
  {A\, e^{-\sigma_{\lambda}N_{H_{2}}^0(\bar{r})} + B} \biggr),
\end{displaymath}
where $A$ and $B$ are two parameters depending only on $N_{H_{2}}^{flat}$ 
and $N_{H_{2}}^{out}$, and $\sigma_{\lambda}$ is again the extinction cross-section at the mid-IR
wavelength $\lambda$. 

For comparison with the profiles derived from the ISOCAM images, 
the thin solid curves in Fig.~\ref{profiles} represent the 
column density profiles of finite-sized sphere models with uniform density 
for $r < R_{flat}$, $\rho \propto r^{-2}$ for $R_{flat} < r < R_{edge}$, 
and surrounded by a medium of uniform column density. 
The dotted  curves plotted in 
Fig.~\ref{profiles} are $N_{H_{2}} \propto \bar{r}^{-1}$ power laws (which correspond to the column density profile of a singular isothermal sphere at 10~K). These models as well as other theoretical models will be discussed in Sect.~4. 
(Note that the convolution with the ISOCAM psf has {\it no} effect on the 
models beyond an angular radius of $\sim$~6$\arcsec$.)

\begin{table*}[ht]
\caption[]{Core parameters derived from the ISOCAM absorption analysis}
\begin{minipage}{16.5cm}
\begin{tabular}{lrrcccccccccc} \hline
Source & $N_{H_{2}}^{flat}$ & $N_{H_{2}}^{peak}$ & $\tau_{max}^{7\mu m}$ & $R_{flat}^{ISO}$& $R_{mid}^{ISO}$ & $R_{edge}^{ISO}$ & $n_{H_{2}}^{flat}$ & $M_{flat}$ & $M_{edge}$ & $m_{1}$ & $m_{2}$ & $\frac{n_{flat}}{n_{out}}$\\
Name & (cm$^{-2}$) &  (cm$^{-2}$) & & (0.01\,pc) & (pc) & (pc) &(cm$^{-3}$) & ($\sm$) & ($\sm$) &  & & \\
 & (1) & (2) & (3) & (4) & (5) & (6) & (7) & (8) & (9) & (10) & (11) & (12)\\
\hline
\hline
L1544 & 5.4$\times$10$^{22}$ & 6.0$\times$10$^{22}$ & 0.7 &1.4 & 0.028 & 0.043 & 4$\times$10$^{5}$ & 0.5 & 1.9 & $-$1.6$^{+0.1}_{-0.2}$ & $-$4.4$^{+0.7}_{-0.1}$ & 20\\
Oph D & 5$\times$10$^{22}$ & 6.6$\times$10$^{22}$ & 0.7 & 1.6 & 0.028 & 0.066 & 3.0$\times$10$^{5}$ & 0.8 & 2.0 & $-$0.8$^{+0.1}_{-0.2}$ & $-$2.8$^{+0.7}_{-2.6}$ & 80 \\
L1709A & 4.5$\times$10$^{22}$ & 5.1$\times$10$^{22}$ & 0.6 & 3.2 & 0.18 & $>$0.32 & 1.2$\times$10$^{5}$ & 2.4 & $>$29 & $-$1.2$^{+0.5}_{-0.4}$ & - & 70\\ 
L1689B & 4.5$\times$10$^{22}$ & 5.8$\times$10$^{22}$ & 0.7 & 3.2 & 0.096 & 0.136 & 1.4$\times$10$^{5}$ & 2.2 & 11 & $-$1.4$^{+0.3}_{-0.2}$ & $-$4.3$^{+1.9}_{-2.9}$ & 50\\
 (South) \\
L1689B & 4.7$\times$10$^{22}$ & 6.0$\times$10$^{22}$ & 0.7 & 4 & 0.28 &$>$0.40 & 1.2$\times$10$^{5}$ & 3.5 & $>$70 & $-1.0^{+0.4}_{-0.8}$ & - & 50\\
 (West)\\
L310 & 3.4$\times$10$^{22}$ & 3.7$\times$10$^{22}$ & 0.5 & 4 & 0.07 & 0.1 & 9$\times$10$^{4}$ & 1.5 & 7.7 & $-$1.5$^{+0.1}_{-0.1}$ & - & $>$10\\
L328 & 4.4$\times$10$^{22}$ & 5.8$\times$10$^{22}$ & 0.8 & 2.5 & 0.03 & 0.12 & 1.8$\times$10$^{5}$ & 1.1 & 6.4 & $-$1.7$^{+0.4}_{-0.7}$ & $-$1.7$^{+0.4}_{-0.7}$ & $>$20\\
L429 & 9.6$\times$10$^{22}$ & 1.2$\times$10$^{23}$ & 1.6 & $\sim$1.2 & 0.35 & $>$0.5 & 6$\times$10$^{5}$ & $\sim$0.4 & $>$60 & $-$0.9$^{+0.5}_{-0.5}$ & - & $>$600 \\
GF5 & 2.7$\times$10$^{22}$ & 3$\times$10$^{22}$ & 0.4 & 2.5 & 0.2 & $>$0.4 & 1$\times$10$^{5}$ & 0.7 & $>$30 & $-$0.9$^{+0.2}_{-0.2}$ & - & $>$140 \\ 
\hline 
\end{tabular}
Notes to Table \ref{coldens}:\\
(1) Column density averaged over the flat inner part of the core of radius $R_{flat}^{ISO}$.\\
(2) {\it Peak} column density read from the ISOCAM profiles of Fig. \ref{profiles}.\\
(3) Opacity at the centre of the core at $\sim$~7\,$\mu$m.\\
(4) $R_{flat}^{ISO}$ is the radius of the flat inner part of the ISOCAM profiles of Fig. \ref{profiles} and is slightly different from $R_{flat}$ (cf. Table \ref{mm}) for L1544, Oph~D and L1689B.\\
(6) Radius of the core edge (when there is one).\\
(7) Average density over the flat inner core region, estimated from a piecewise power-law model fitting the profile (see text).\\
(8) Integrated core mass within a circular area of radius $R_{flat}^{ISO}$.\\
(9) Integrated core mass within a circular area of radius $R_{edge}^{ISO}$. \\
(10) Index of the power-law fitting the column density profile between $R_{flat}^{ISO}$ and $R_{mid}^{ISO}$ (Column 4).\\
(11) Index of the power-law fitting the column density profile between $R_{mid}^{ISO}$ (Column 4) and $R_{edge}^{ISO}$.\\
(12) Density contrast between the centre and the edge of the core. For Oph~D, L1689B and L1709A, the value of $n_{out}$ was taken from the large scale $^{13}$CO observations of Loren (1989a), and for L1544, it comes from 
Butner et al. (1995). 
As no estimate of $n_{out}$ is available for the remaining cores in the literature, a lower limit for the density contrast was estimated from the model described in Section~\ref{BE}.\\
\end{minipage}
 \label{coldens}
 \end{table*}

\subsubsection{Results}

It can be seen from Fig.~\ref{profiles} that
all the column density profiles 
present a flattening in their inner regions - though this flattening is only marginal in the case of L429.
On the other hand, the slopes and shapes of the outer parts of the
profiles appear to vary from core to core and may be separated into
two broad categories. For four of the cores
(GF5, L429, L1709A and the Western part of L1689B), the profile
beyond the flat inner part follows a near power law with projected radius $\bar{r}$, i.e.
$N_{H_{2}} \propto \bar{r}^{-1}$, and does not show any clear steepening up to
the boundary of the image.
By contrast, for the other cores (L328, Oph D, L310, L1544 and the Southern side of L1689B), the profile steepens
beyond some radius $R_{mid}$, until it merges with
the ambient molecular cloud.
Three cores in our sample present ``edges'', defined
as regions where the column density profile drops significantly steeper than a
$N_{H_{2}} \propto \bar{r}^{-1}$ power law. 
Globally, we may divide each column density profile into four parts:
a flat inner region of radius $R_{flat}^{ISO}$, followed by
a near power-law portion of index $m_{1}$ up to a radius $R_{mid}^{ISO}$, a region
with a steep slope $m_{2}$ from $R_{mid}^{ISO}$ to $R_{edge}^{ISO}$ that marks the
core ``edge'', and finally a relatively flat plateau 
once the column density has reached the average value it has in
the ambient molecular cloud. Depending on the cases, one or two parts
may be absent.

The values of the core opacities at 7~$\mu$m, the column
densities, as well as the masses, average slopes, and density contrast deduced from the ISOCAM profiles of Fig.~\ref{profiles} are given in 
Table~\ref{coldens}.
The peak column density, $N^{peak}_{H_{2}}$, was read on the profiles
shown in Fig.~\ref{profiles}. (In most cases, $N^{peak}_{H_{2}} > N^{flat}_{H_{2}}$ because the central plateau is not strictly flat).
The average slopes $m_{1}$ and $m_{2}$ of the column density profile between 
$R_{flat}^{ISO}$ and $R_{mid}^{ISO}$, and between $R_{mid}^{ISO}$ and $R_{edge}^{ISO}$,
respectively, were derived by fitting segments of straight lines
to the data. 
The maximum values of $m_{1}$ and $m_{2}$ were obtained by assuming the maximum possible value of $N^{flat}_{H_{2}}$ together with the minimum possible value of $N_{H_{2}}^{out}$ (taking into account the factor of $\sim$\,2 uncertainty), while the minimum values of $m_{1}$ and $m_{2}$ were obtained by assuming the minimum possible value of $N^{flat}_{H_{2}}$ together with the maximum possible value of $N_{H_{2}}^{out}$.

\begin{figure*}
\psfig{file=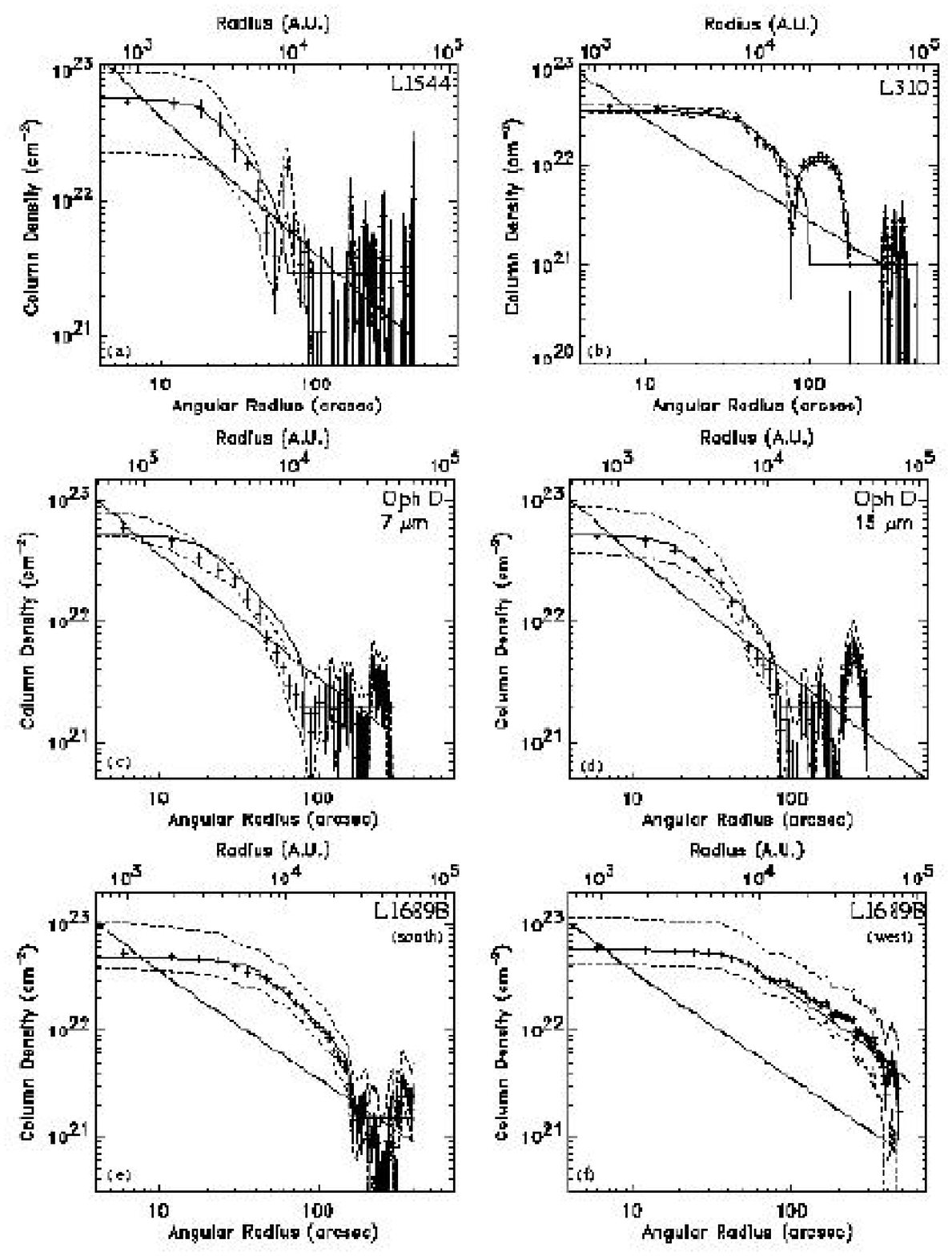,width=17cm}
\caption{Column density profiles of the target pre-stellar cores (crosses),  
as well as global error bars (dashed curves), derived from the ISOCAM images 
and adjusted so as to match the values obtained in 
the flat central region with 1.3~mm continuum data and in the outer region 
with C$^{18}$O(1--0) line observations (see \S ~\ref{modelling}). 
Theoretical models are plotted for comparison (see \S ~\ref{discussion}).}
\label{profiles}
\end{figure*}
\setcounter{figure}{4}
\begin{figure*}
\psfig{file=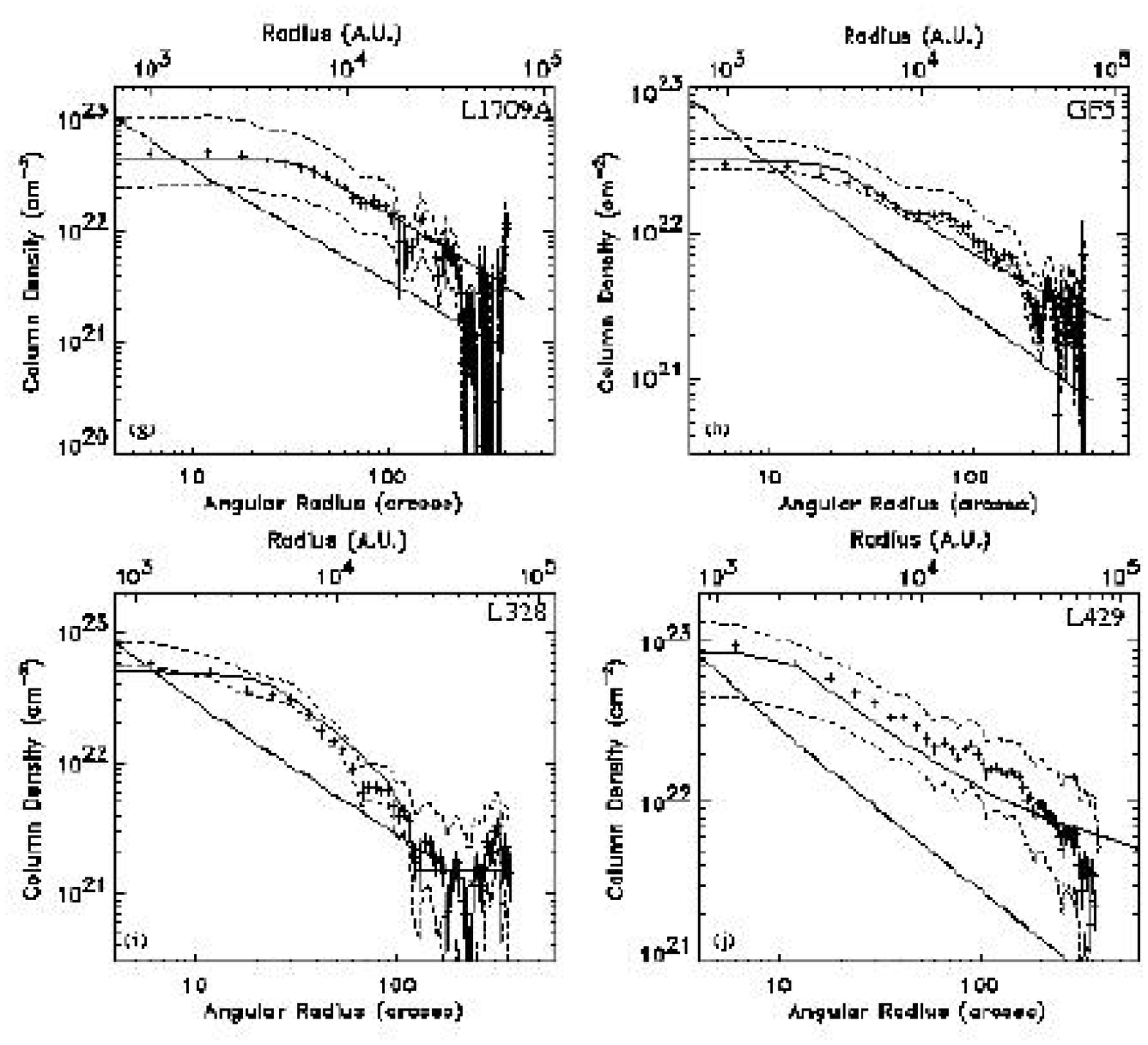,width=19cm}
\caption{(cont'd) The dotted line is an $N_{H_{2}} \propto \bar{r}^{-1}$ power-law (corresponding to the profile of a singular isothermal sphere at T=10~K), convolved with the ISOCAM psf. The thin solid line corresponds to the column density profile of a finite-sized sphere model with uniform density for $r < R_{flat}$, $\rho \propto r^{-2}$ for $R_{flat} < r < R_{edge}$, 
and surrounded by a medium of uniform column density.} 
\end{figure*}

\subsection {Uncertainties, difficulties, limitations}

\label{uncert}

\subsubsection {Dust extinction curve in the mid-IR}

Among the parameters involved in the relationship between the mid-IR
intensity and the H$_{2}$
column density is the dust extinction coefficient $\sigma_{\lambda}$. 
This paragraph focuses on the influence of $\sigma_{\lambda}$ on the derived column density profiles.
The uncertainty on $\sigma_{\lambda}$ stems from two main factors. 
First, the ISOCAM filters encompass 
a certain bandwidth over which the value of $\sigma_{\lambda}$ varies. 
In our study, we used for $\sigma_{\lambda}$ the value for the wavelength 
at the centre of the filter (ie at 6.75 $\mu$m for the LW2 filter and 7.75 
$\mu$m for the LW6 filter). Considering the rapid variations (steep increase) 
of $\sigma_{\lambda}$ towards 8~$\mu$m (eg. Draine \& Lee 1984), the values
we adopted in \S ~\ref{modelling} above could be underestimated by up to 
a factor of $\sim$~2. Second, the dust properties
in dense cores are known to depart from those calculated by Draine \& Lee (1984) for the diffuse
interstellar medium (Kr\"ugel \& Siebenmorgen 1994, see in particular their Fig.
12). In environments
like dense cores, in which temperatures can be as low as $\sim$10~K and
densities as high as
$\sim 10^{5}$-$10^{6}$ cm$^{-3}$, coagulation of particles, fluffiness and the presence of ice mantles
on the grains can change the value of the dust extinction coefficient, depending on the nature
of the grains (amorphous carbon or coated silicate grains). By using the Draine \& Lee value,
we could thus be underestimating the value of the extinction coefficient by a factor of $\sim$~2.

To check how the column density profile was affected by the value of $\sigma_{\lambda}$, we
calculated it for different values of $\sigma_{\lambda}$. As the innermost parts of the core are
denser and more shielded from the interstellar radiation field (and thus colder), they are more
likely to be affected by grain coagulation and forming of ice mantles, 
resulting in a higher extinction coefficient than in the low-density outer
parts. We therefore determined the column density
profiles supposing a linear increase of $\sigma_{\lambda}$ from 1.35$\times 10^{-23}$cm$^{2}$ on the edges
of the core 
to 2.7$\times 10^{-23}$cm$^{2}$ (which is
twice the previous value) at the centre of the core. 
We also varied the value of
$\sigma_{\lambda}$ by factors of 2 and 3 to check the influence of a high value of $\sigma_{\lambda}$ on the shapes of the profiles.

The results of this study show that if $\sigma_{\lambda}$ varies within a factor of $\sim $~2--3 the column density profiles change only marginally. 
The same holds true if we assume a linear gradient in $\sigma_{\lambda}$ 
or in Log($\sigma_{\lambda}$) from the centre to the edge of the cloud 
(so that in the end
$\sigma_{\lambda}$ varies within a factor of 2).
To summarize, 
our conclusions regarding the slopes of the column density profiles are not 
sensitive to the existing uncertainties on $\sigma_{\lambda}$.

\subsubsection {Simulations of non uniform background and foreground intensities}
\label{nonuni}


\begin{figure*}[htbp]
\psfig{file=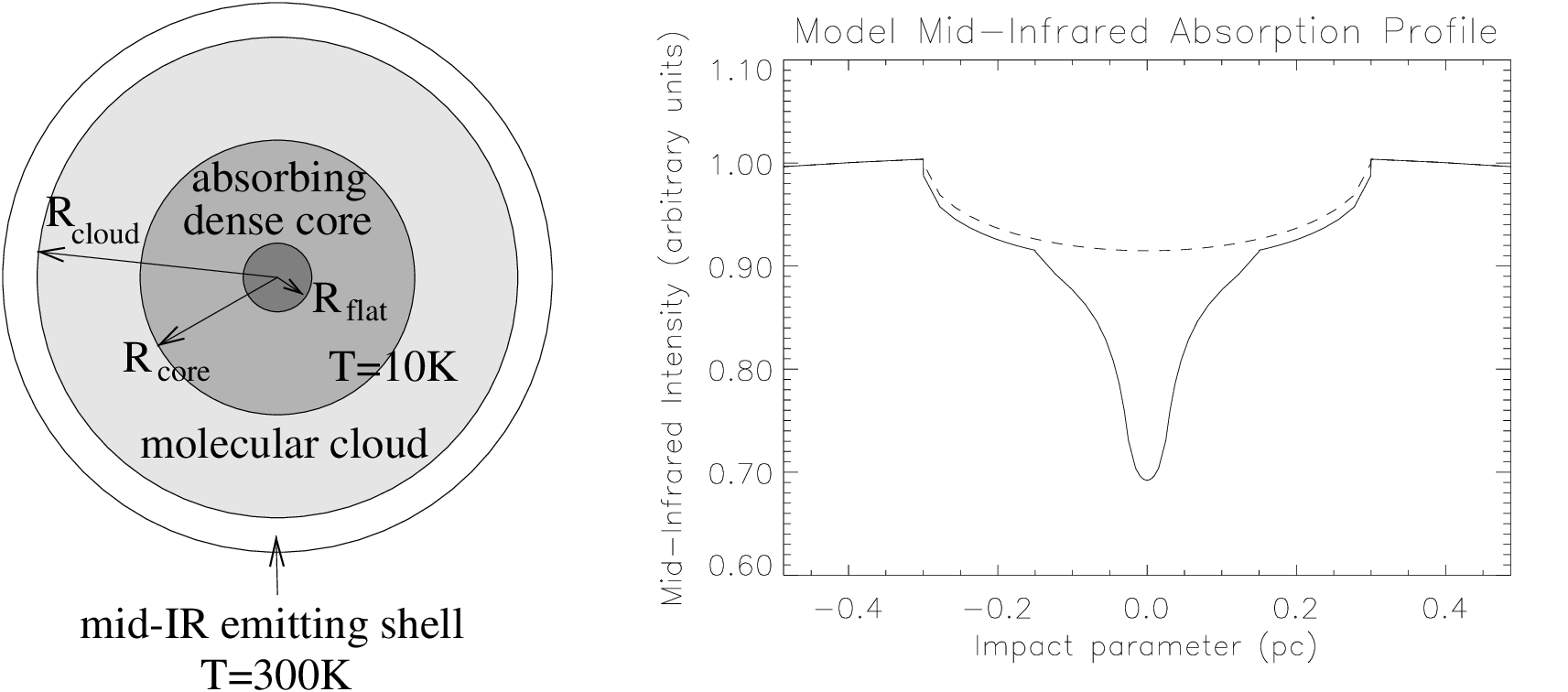,width=18cm,angle=270}
\caption{Schematic representation of the simple model cloud/core system 
discussed in the text (left panel). The right panel shows the simulated 
mid-IR intensity profile (solid curve) emerging from such a model. 
The dashed curve represents what the intensity profile would be without 
the absorbing dense core/cloud. 
The illustrative case shown has $R_{flat} = 0.025$~pc, $R_{core} = 0.15$~pc,
and $R_{cloud} = 0.3$~pc. Note some similarity with the observed profiles 
of Fig.~3.
}
\label{absmodel}
\end{figure*}

To check the influence of our assumption that the foreground and background intensities are spatially 
uniform, we carried out simulations of the total infrared intensity 
measured by the camera (and of
the column density inferred), if the absorbing dense core is 
embedded in an ambient cloud of uniform column density, itself bounded 
by a mid-IR emitting shell (cf. Fig.~6 -- left panel).
We considered a spherically symmetric core model with 
uniform density for $r < R_{flat}$, $\rho \propto r^{-2}$ for 
$R_{flat} < r < R_{core}$ (corresponding to a $N \propto \bar{r}^{-1}$ 
column density profile), and uniform temperature 
$T_{in} \sim 10$~K throughout. (This is similar to the simple model 
compared to the data in Fig.~5 and discussed in Sect.~4.1 below.) 
The parent cloud (corresponding to radii $R_{core} < r < R_{cloud}$) is also
assumed to be at $T_{in} \sim 10$~K. By contrast, the outer shell 
(beyond $R_{cloud}$) is taken to be much warmer at $T_{out} \sim 300$~K,
so that its maximum emission is in the mid-IR.  
Thus, we have a thin outer region emitting mid-IR radiation and a cold, 
dense inner cloud/core only absorbing in the mid-IR. 
Such an idealized model of emission by an outer cloud in thermal equilibrium 
at $T_{out} \sim 300$~K is clearly not realistic 
and only meant to be illustrative of a background medium 
emitting in the mid-IR. 
Since it is the thin 
spherical shell beyond $R_{cloud}$ that produces the background and 
foreground, $I_{back}$ and $I_{fore}$ are {\it not} uniform 
in this model but present `limb-brightened' peaks at 
$\bar{r} = R_{cloud}$ (see Fig.~\ref{absmodel} -- right panel). 

When $R_{core}$ is significantly smaller than $R_{cloud}$, our simulations 
show that $I_{back}$ and $I_{fore}$ vary only slowly over the spatial 
extent of the absorbing dense core (see the example of Fig.~6). In such a 
situation, the column density profile derived from the simulated 
mid-IR intensity of the model in the same way as for the ISOCAM data 
is almost indistinguishable from the original model profile.
On the other hand, if $R_{core} \sim R_{cloud}$ so that the peaks 
in $I_{back}$ and $I_{fore}$ occur immediately outside the core, then the 
profile derived under the assumption that $I_{back}$ and $I_{fore}$ are 
spatially uniform departs markedly from the true column density profile
as the temperature discontinuity is approached for $\bar{r}\sim R_{cloud}$. In particular,
the derived profile drops sharply near $R_{core}$, i.e., displays a sharp 
edge, even when the warm outer shell has the same $\rho \propto r^{-2}$ density
profile as the cold inner cloud core (so that there is no physical edge 
in the density structure but only a discontinuity in temperature).
This shows that our method of converting intensities into column densities
enables us to detect the existence of a discontinuity, but does not 
necessarily allow us to conclude about its nature.

\begin{table*}[ht]
\caption[]{Values of reduced $\chi^{2}$ for the various theoretical models}
\begin{minipage}{12cm}
\begin{tabular}{lccccc} \hline
Source & Parameterized & Logotrope & best fitting SIS$^{1}$& BM95 & CM95 \\
& Sphere & & & & \\
\hline \hline
L1544 & 5.9 & 8.2 & 25.8 & 11.3 & 5.2 \\
Oph D & 1.3 & 13.4 & 23.0 & 19.9 & 1.3 \\
L1709A & 3.0 & 5.5 & 8.8 & 3.0 & - \\
L1689B s & 1.0 & 9.8 & 29.4 & 1.2 & 17.7 \\
L1689B w & 3.2 & 47.2 & 241.7 & 3.7 & 21.1 \\
L328 & 4.3 & 30.8 & 19.5 & 49.7 & 6.8 \\
L429 & 19.2 & 15.4 & 14.3 & 5.7 & - \\
GF5 & 8.0 & 4.4 & 54.0& 3.3 & - \\
\hline 
\end{tabular}
Notes:\\
The $\chi^2$ calculations were restricted to radii smaller than $R_{edge}$ (Table \ref{coldens}) for cores with an edge.\\
$^{1}$~The SIS model considered in this Table is the one that fits best the near power-law portion of the
column density profile. It corresponds to gas temperatures higher than 10~K.
\end{minipage}
\label{chi2}
\end{table*}
In practice, however, it seems reasonable to assume that we are in the 
situation of Fig.~6 and that $I_{back}$ and $I_{fore}$ have 
nearly uniform values over distance scales of several tenths of parsecs. 
The $\rho$~Ophiuchi cloud complex may be used as an example. 
In this case, there is convincing evidence from the IRAS study of 
Bernard et al. (1993) that the mid-IR background emission originates 
from photo-excited PAH-like molecules distributed in a sort of shell 
corresponding to the outer layers of the large-scale $^{13}$CO molecular 
cloud mapped by Loren (1989).
This geometry is very reminiscent of Fig.~6 with $R_{cloud}$ on the order of 
{\it a few pc}, i.e., much bigger than 
the rather small ($R_{core} \sim 0.1$~pc) cores we are studying here 
(see, e.g., Table~\ref{coldens}). This means that the cores
are embedded deep inside the layers responsible for the mid-IR emission.
Another piece of evidence comes from the ISOCAM images themselves which do not show strong intensity fluctuations on small scales.

\section{Discussion: Comparison with theoretical models of core structure}

\label{discussion}

As pointed out in Sect.~\ref{coldensprof}, the radial column density profiles derived for the 
pre-stellar cores of our ISOCAM absorption sample (see Fig. \ref{profiles}) are generally characterized 
by four different regimes: a) a flat inner region where the column density 
gradient is much flatter than $N_{H_2} \propto \bar{r}^{-1} $, b) a region roughly 
consistent with $N_{H_2} \propto \bar{r}^{-1} $, c) an edge where the column 
density gradient steepens quickly with distance from core center and gets 
as steep as $N_{H_2} \propto \bar{r}^{-2} $ or even steeper (see Table~\ref{coldens}),
until d) the end 
of the dense core is reached and $N_{H_2}$ fluctuates about the mean value 
in the parent cloud (i.e., typically 
$N_{out} \sim 2 \times 10^{21}$~cm$^{-2}$). 

In the simple case of an infinite spheroidal core with a power-law 
density structure, the radial density
gradient, $\rho \propto r^{-p} $, is related to the column density gradient,
$N_{H_2} \propto \bar{r}^{-m}, $ by $ p = m + 1 $ (e.g. Adams 1991, 
Yun \& Clemens 1991). If alternatively, the core radial density structure is 
not represented by a single power law but by piecewise power laws, then 
the preceding relation between $ p $ and  $ m $ 
is no longer strictly exact in the transition regions 
due to convergence effects. However, these effects can be quantified in 
specific cases.
For instance, in the case of a sphere of uniform density 
at both small ($r < r_0$) and large ($r > r_1$) radii, and having 
$\rho \propto r^{-2}$ between $r_0$ and $r_1$, the column density profile 
gets somewhat steeper than $N_{H_2} \propto \bar{r}^{-1}$ between $r_0$ and $r_1$, 
but remains always shallower than $N_{H_2} \propto \bar{r}^{-1.4}$. For a similar 
sphere with $\rho \propto r^{-3}$ instead of $\rho \propto r^{-2}$
between $r_0$ and $r_1$, the column 
density profile is always shallower than $N_{H_2} \propto \bar{r}^{-2.2}$. We 
conclude that, if the column density profile 
is observed to become steeper than $N_{H_2} \propto \bar{r}^{-2.2}$, then 
we may safely infer that the underlying density profile becomes steeper 
than $\rho \propto r^{-3}$ at some neighboring radius. 
Note that this conclusion 
remains essentially valid even in the extreme case of a {\it finite-sized} 
sphere where the density drops to $0$ beyond a radius $R_{out}$ 
(cf. Arquilla \& Goldsmith 1985, Yun \& Clemens 1991). In such a  case, both the density and the column 
density gradient become effectively {\it infinite} at $R_{out}$. 
Although the simple relation $ p = m + 1 $ would yield too large values of $p$ 
in a narrow range of radii $r \simlt R_{out}$, the basic conclusion that 
$p$ becomes infinitely large close to $R_{out}$ would still be correct.

Characteristics a) and b) above emphasize the fact that pre-stellar
cores flatten out near their centres and are consistent with previous
(sub)millimeter continuum results (Ward-Thompson et al. 1994; AWM96; WMA99).
Characteristic c), implying an outer density profile steeper than 
$\rho \propto r^{-3}
$ for three or four of our cores (see Table \ref{coldens}) is new and has interesting
implications. 

In the following subsections, we discuss the physical 
meaning of the inferred profiles by comparing them with the predictions 
of various theoretical models proposed for pre-stellar cores. For each model column density profile, $N_{Mod}(\bar{r})$, we calculated a value of reduced 
$\chi ^2$ according to the formula:

\begin{equation}
\chi ^2 = \frac{1}{N_{radii}} \, 
\sum \frac{[I(\bar{r}_i) - I_{back}e^{-\sigma_{MIR}N_{Mod}(\bar{r}_i)} - I_{fore}]^2}
{\epsilon_i ^2},
\end{equation}
where the sum is over the sampled radii, and $\epsilon_i$ is the 
observational rms uncertainty estimated by calculating the 
dispersion of the mid-IR intensities that were averaged at radius $r_i$ in 
the image, divided by the square root of the number of points at $r_i$ 
(cf. Sect.~\ref{coldensprof}). These
reduced $\chi ^2$ values which provide an estimate of the ``goodness'' of the various
models are listed in Table~\ref{chi2}.

\subsection{Comparison with simple parameterized spherical models}

\label{BE}

Most of the column density profiles obtained in Sect.~\ref{coldensprof} can be reasonably 
well fitted by finite-sized sphere models with piecewise power-law density 
gradients.
The model column density profiles shown as thin solid curves in Fig.~\ref{profiles} 
were integrated for truncated spheres of uniform central density $n_{c}$ 
up to a radius $R_{flat}$, and of density decreasing as 
$\rho \propto r^{-p}$ between 
$R_{flat}$ and an outer truncation radius $R_{edge}$.
Beyond $R_{edge}$, the total ambient column density, $N_{H_2}$,  
is supposed to be uniform. 
These simple parameterized models provide good approximations to the profiles 
of polytropic gas spheres (e.g. Chandrasekhar 1967, Chi\`eze 1987, 
McKee \& Holliman 1999) and more general equilibrium models 
for externally-heated, thermally-supported self-gravitating spheroids 
(Falgarone \& Puget 1985, Chi\`eze \& Pineau des For\^ets 1987).
The special case 
$p = 2$ is the most relevant here since the power-law portion of the observed
column density profiles is very close to $N_{H_2} \propto \bar{r}^{-1} $ 
(see above). In this case, the models mimic the column density profiles 
of pressure-bounded Bonnor-Ebert spheres (Bonnor 1956, Ebert 1955). 
These correspond to the spherical solutions of the equation of hydrostatic 
equilibrium for isothermal self-gravitating cores confined 
by external pressure. They have been considered as plausible starting points 
for the gravitational collapse of protostellar cores in quiescent clouds 
(e.g. Shu 1977, Foster \& Chevalier 1993).

The parameters $n_{c}$, $R_{flat}$, and $R_{edge}$, of the `best-fit' models for $p =2$ are 
listed in Table \ref{coldens} for each of the cores with an edge. 
These Bonnor-Ebert-like models generally yield the lowest values 
of reduced $\chi ^2$ among the models we have investigated (cf. Table \ref{chi2}). We stress, however, 
that since most of our cores exhibit non-spherical 
morphologies they can at best be roughly approximated by such models. 
Furthermore, the density contrasts ($\sim $~10--80 -- see Table~\ref{coldens}) 
that we infer from center to edge for our cores 
are generally larger than the maximum contrast 
of $\sim 14$ for stable Bonnor-Ebert spheres. If thermal pressure were the 
only source of support against gravity, most of the cores in our sample 
should thus be gravitationally unstable. However, since the 
lifetimes estimated for similar starless cores are 
$\simgt 10^6$~yr (e.g. Lee \& Myers 1999, Jessop \& Ward-Thompson 2000), 
i.e., a factor of $\simgt 10$ larger than free-fall timescales given the 
central densities estimated in Table~\ref{coldens}, 
it is likely that the cores experience extra support from nonthermal 
pressure forces, especially in their outer regions. 
In this respect, the `non-isentropic' multi-pressure 
polytropes recently proposed by McKee \& Holliman (1999), which include 
a stabilizing magnetic pressure component allowing large density contrasts, 
offer an interesting solution (although they do not account for the 
aspherical core morphologies).

\subsection{Comparison with logotropic spheres}

McLaughlin \& Pudritz (1996, 1997) have advocated  
a special non-isothermal equation of state for star-forming molecular clouds 
of the `logotropic' form $P/P_c = 1 + A\, $ln$(\rho/\rho_c)$, where 
$P_c$ and $\rho_c$ denote the central values of the pressure and density, 
respectively, and $A$ is a constant whose realistic value is 
$A \approx 0.2$. Assuming this equation of state, 
McLaughlin \& Pudritz show that the hydrostatic equilibrium solutions 
for pressure-truncated self-gravitating spheres are analogous 
to Bonnor-Ebert isothermal spheres with flat inner regions  
but with $\rho \propto r^{-1}$ (rather than $\rho \propto r^{-2}$) 
density gradients in their outer parts. 
Stable logotropic equilibria exist for much larger density contrasts 
(up to $\simgt 100$) than in the isothermal case.

%
We compared the column density
profiles inferred from our ISOCAM data with the profiles of finite-sized 
logotropes at a gas temperature\footnote{The results of this comparison would remain unchanged if we
used T=12.5~K, the dust temperature adopted for our pre-stellar cores -- see 
Sect.~3.3.1.} of T=10~K, 
bounded by a surface pressure $P_{s}$. We investigated various values 
of $P_{s}$ and of the center-to-edge pressure contrast $P_{c}/P_{s}$ (see 
Table~\ref{logo}).
\begin{table}[t]
\caption[]{Characteristics of the logotropes compared to the data.}
\begin{tabular}{lccccc} \hline
Name & $P_{s}/k$ & $P_{c}/P_{s}$ & $n_{c}$ & $R_{flat}$ & $R_{edge}$ \\
& cm$^{-3}$ K & & cm$^{-3}$ & 10$^{-2}$ pc & pc \\
\hline
\hline
L1544 & 2.3$\times$10$^{6}$ & 2.68 & 6$\times$10$^{5}$ & 0.3 & 0.07 \\
Oph D & 2.3$\times$10$^{6}$ & 2.68 & 6$\times$10$^{5}$ & 0.3 & 0.07\\
L1709A & 2$\times$10$^{5}$ & 20 &2$\times$10$^{5}$ & 0.35 & 0.4\\
L1689B s& 1.4$\times$10$^{6}$  & 4.16 & 6$\times$10$^{5}$ & 0.3 & 0.13\\
L310 & 2.3$\times$10$^{6}$ & 2.68 & 5$\times$10$^{5}$ & 0.3 & 0.07 \\
L328 &  7$\times$10$^{5}$ & 4.16 & 2.8$\times$10$^{5}$ & 0.4 & 0.19\\
L429 & 2.5$\times$10$^{5}$ & 20 &  5$\times$10$^{5}$ & 0.3 & 0.33 \\
GF5 & 4.9$\times$10$^{5}$ & 4.16 & 2$\times$10$^{5}$ & 0.5 & 0.22 \\
\hline
\label{logo}
\end{tabular}
\end{table}
Once these parameters are fixed, the logotrope is entirely defined 
and the central density, $n_{c}$, the radius of the flat inner region, $R_{flat}$, and the truncation radius of the model, $R_{edge}$, 
can all be determined (see Table 1 of McLaughlin \& Pudritz 1997 
and Table~\ref{logo}). 

\begin{figure*}
\psfig{file=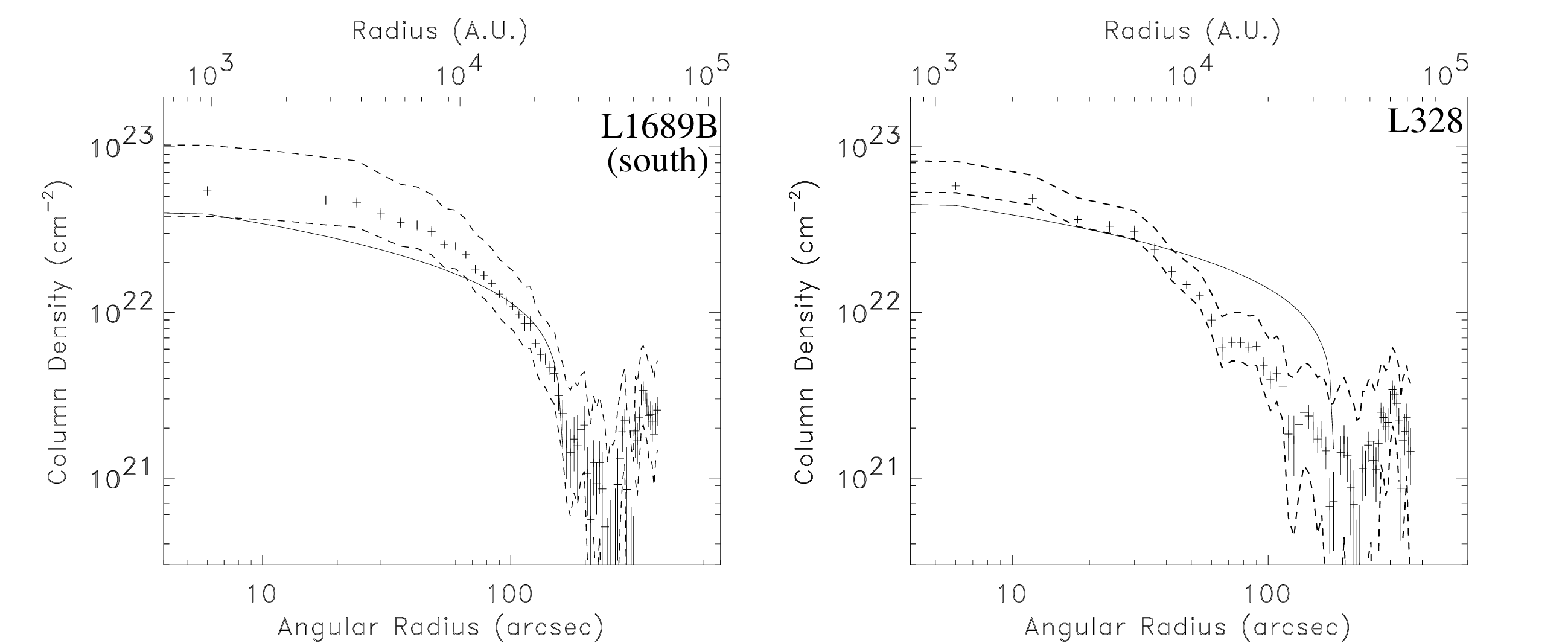,width=19cm,angle=-90}
\caption{Comparison of the logotropic model of McLaughlin \& Pudritz (1996) 
with the column density profiles of L1689B (south) and L328. The column 
density profile of this model is too shallow to fit the observations. }
\label{logomodel}
\end{figure*}

Figure~\ref{logomodel} 
shows (as solid curves) the best-fitting logotrope (to which we added a
uniform baseline in column density) superimposed on the graphs of the experimental column
density profiles of L1689B (South sector) and L328.

From examination of Fig.~\ref{logomodel}, it appears that the logotropic model 
cannot match the observed core profiles. 
The disagreement is due to a difference 
in the overall shape/slope of the column density profiles beyond the flat 
inner region: while the observed profiles approach $N_{H_2}(\bar{r}) \propto \bar{r}^{-1} $, the logotrope is substantially flatter than this  
[with approximately $N_{H_2}(\bar{r}) \propto \,$ln$(\frac{2\, R_{edge}}{\bar{r}}) $ for 
$\bar{r} << R_{edge}$]. Furthermore, since all realistic logotropes must be 
truncated at $R_{edge} \simlt 0.5$~pc to remain stable against radial 
perturbations (see Table~1 of MacLaughlin \& Pudritz 1997),
they cannot account for ``edgeless'' cores such as L429, GF5, and L1709A. 
More quantitatively, the reduced $\chi ^2$ values for the `best-fit' 
logotropes are always larger than $\sim 5$, and it is always possible to 
find a better model with a substantially lower value of $\chi ^2$ 
(see Table \ref{chi2}).



\subsection{Comparison with self-similar, singular models}

\label{singular}
The most extreme versions of the isothermal and logotropic models considered 
above are the singular isothermal sphere (SIS -- e.g. Shu 1977) 
and the singular logotrope (e.g. McLaughlin \& Pudritz 1997). These are 
hydrostatic self-gravitating spheres with power-law density profiles such 
as $\rho(r) = (a^2/2\pi G)r^{-2} $ and $\rho(r) = (A P_c/2\pi G)^{1/2}r^{-1}$,
respectively, where $a$ is the effective isothermal sound speed. Although 
the singular models 
are well into the unstable regime of Bonnor-Ebert and logotropic spheres, 
they present the great advantage of allowing semi-analytical similarity 
solutions for the collapse (Shu 1977, McLaughlin \& Pudritz 1997). 
Such singular models could a priori be representative of the most advanced
pre-stellar cores, i.e., those on the verge of collapse and/or protostar 
formation. 
In particular, the SIS (or slight variants of it) 
provides the collapse initial conditions in the widely used 
`standard' paradigm of isolated star formation (e.g. Shu et al. 1987). 

These singular models, and especially the SIS, are much too steep at small 
radii to match the flat inner region characterizing the observed 
pre-stellar profiles (cf. Fig.~\ref{profiles}; see also Ward-Thompson 
et al. 1994, 1999, and AWM96).  
Furthermore, as such, these models are scale-free 
and do not account for the presence of edges either. The combination of these 
two qualitative differences with the observed profiles imply fairly large 
values of $\chi ^2$ (see Table~\ref{chi2}). 

The fact that the observed profiles are much flatter than the singular models 
at small radii could imply either that the cores of our sample are not 
sufficiently evolved to be described by such models (cf. Li \& Shu 1996), 
or that the collapse initial conditions are non-singular. Since there is 
already evidence for significant inward motions in (at least some of) our cores (see 
Sect.~4.4 and Fig.~10 below), we favor the latter interpretation.

It is also interesting to note that, even when a large part of the observed 
column density profile is compatible with a $N_{H_{2}} \propto \bar{r}^{-1}$ 
power law (e.g. L429, L1709A), the profile of a SIS at T=10~K 
falls significantly below the actual core profile. In order for the SIS 
profile to match the power-law portion of the observed pre-stellar profiles, 
the effective sound speed must be typically twice as large as the isothermal 
sound speed, ($a \approx 0.2$~km~s$^{-1}$ for T=10~K). 
This corresponds to an overdensity factor of $\sim$~4 
compared to a SIS at 10~K, which again suggests that
there is an additional force (e.g. magnetic or turbulent) supporting the cores
against collapse. In the case of additional turbulent support, however, 
we can exclude the models with thermal and nonthermal 
(TNT) support proposed by Myers \& Fuller (1992), on similar grounds as above, 
i.e., these models are also singular and do not account for the inner 
flattening of the cores.

\subsection{Comparison with magnetic ambipolar diffusion models of cloud cores}

\begin{figure*}
\psfig{file=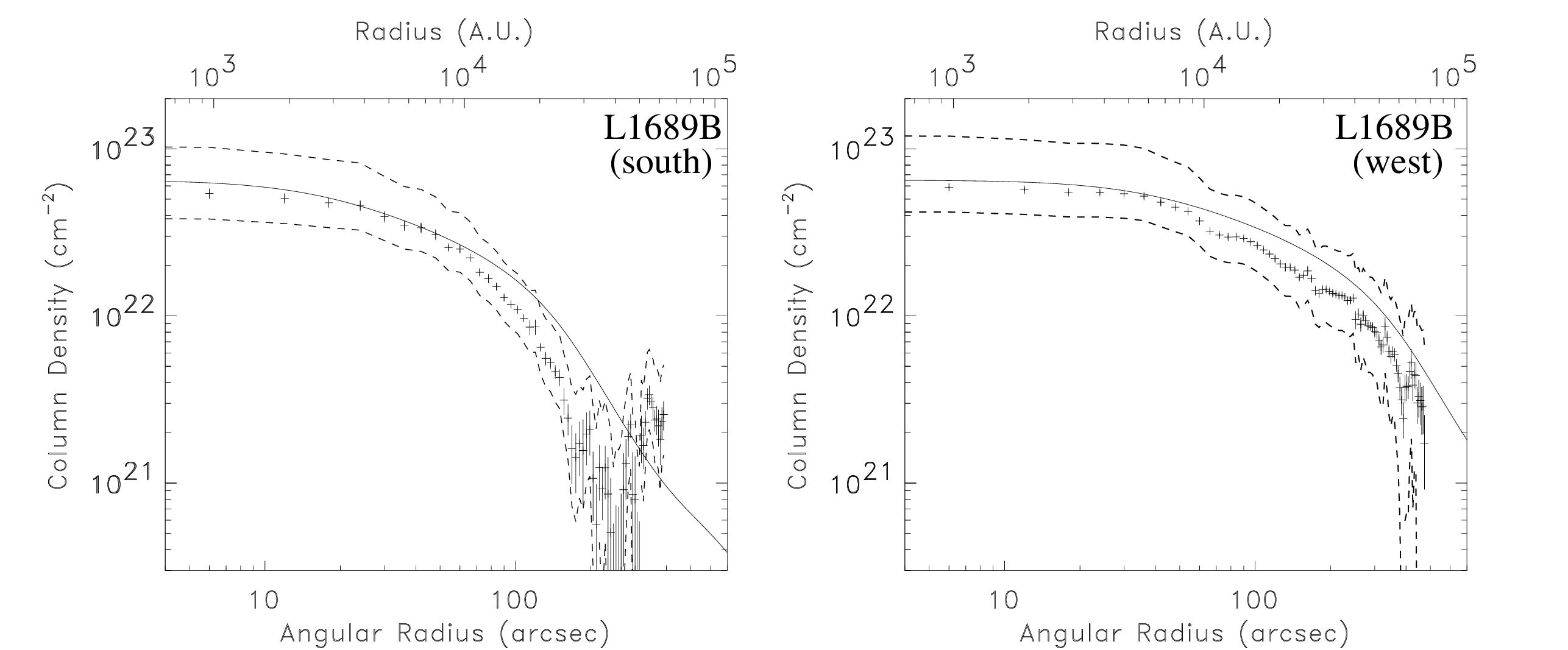,width=19cm,angle=-90}
\caption{Comparison of a magnetically supported core model from BM95 with 
the south and west column density profiles of L1689B. 
Parameters of the magnetic model are: initial central mass-to-flux ratio 
$\mu_{c0} = (M/\Phi)_{c0} = 0.1$ (in units of the critical value $(M/\Phi)_{crit}^{cent}$),
initial central density $n_{c,0}=5.1\times10^{3}$cm$^{-3}$ (see also Table \ref{magn})}
\label{basumodel}
\end{figure*}

A natural way of accounting for flat inner density gradients, high density 
contrasts, nonthermal support, and sharp edges is to consider models of 
magnetically-supported cores undergoing ambipolar diffusion 
(e.g. Ciolek \& Mouschovias 1994, Basu \& Mouschovias 1995 -- hereafter BM95).
According to these models, two phases can be distinguished in the evolution 
of a pre-stellar core. The first phase is a long, quasi-static 
contraction through ambipolar diffusion. Along the magnetic field 
lines, thermal pressure balances gravitational forces while in the 
perpendicular directions, the ions are retained by the magnetic field and 
slow down the process of contraction. During this magnetically 
{\it subcritical} phase, the central mass-to-flux ratio $(M/\Phi)_{cent}$ 
increases until it reaches the critical value 
$(M/\Phi)^{crit}_{cent} = (1/2\pi )G^{-1/2}$
(e.g. Mouschovias 1995, BM95). 
At this point, the inner central core becomes magnetically {\it supercritical} 
and the evolution speeds up. The supercritical core collapses dynamically 
inward, while the outer (subcritical) envelope is still efficiently supported 
by the magnetic field and remains essentially ``held in place''. As a result, 
a (very) steep density profile or `edge' develops at the boundary of the 
supercritical core. The magnetic forces 
induce a disk-like 
geometry for the core, with the polar axis lying along the field lines. 
The volume density of the core is nearly uniform inside a roughly spherical 
central region whose size corresponds to the instantaneous Jeans length. 
Outside this central region, the density declines as $\rho(r) \propto r^{-s}$,  
where the power-law index $s$ varies typically between $\sim 1.5$ 
and $\sim 4.5$ depending on radius and on model parameters such as the 
initial mass-to-flux ratio (see, e.g., Fig.~8 of BM95). 

The column density profiles inferred from our data are qualitatively 
consistent with the predictions of these magnetic models, shortly after 
the formation of the magnetically supercritical core.
%
%
%
%
%
%

In particular, we have quantitatively compared our (South and West) radial 
profiles for L1689B to a range of theoretical profiles at various evolutionary
times calculated by BM95 (cf. Fig.~\ref{basumodel}). 
We find a good agreement between the data and the (rotating) core 
model~7 of BM95 near time $t_{2}$, i.e., the time at which the
central density has increased by a factor of $10^2$ (and the central 
column density by a factor 10) compared to its initial value. This is close to the start of
dynamical collapse in the model. 
The model profiles plotted in Fig. \ref{basumodel} correspond 
to a central number density of $5.1 \times 10^{5}$~cm$^{-3}$ and a temperature of 10~K. 
The Southern and Western profiles of L1689B can both be fitted with the same magnetic model if we assume that L1689B has
the thin disk geometry defined by BM95 and is inclined by an angle 
$i \sim 65\degr$ with respect to the line of sight. Thus, 
the intrinsic column density of the core along the magnetic axis, 
$N_{H_2}^\bot$, is a factor $cos\, i \sim 1/2$ smaller than 
the observed (line-of-sight) column density, i.e., 
$N_{H_2}^\bot \sim 2.3 \times 10^{22}$~cm$^{-2}$ at core centre in this case. Furthermore, in the thin disk approximation, the apparent minor axis radius of the core (in the north-south direction -- see Fig.~\ref{absorption}h) 
should also be a factor $cos\, i \sim 1/2$ smaller than the major axis radius 
(in the east-west direction). 
According to this comparison, L1689B should have just entered the phase 
of dynamical contraction following the formation of the supercritical core.
Given that the edge must be situated just beyond 
$R_{edge} \sim 0.28$~pc~$\sim 5.7 \times 10^4$~AU (cf. Table~\ref{coldens}),
we estimate that the radius of the supercritical core should be 
about $0.03$~pc~$\sim 6000$~AU (i.e., about 10 times smaller than $R_{edge}$ according to model 7 of BM95), 
which is only slightly larger than the radius of the flat inner region 
($\sim 4000$~AU).

An independent argument supporting the claim that L1689B is dynamically 
\begin{figure}[!hb]
\psfig{file=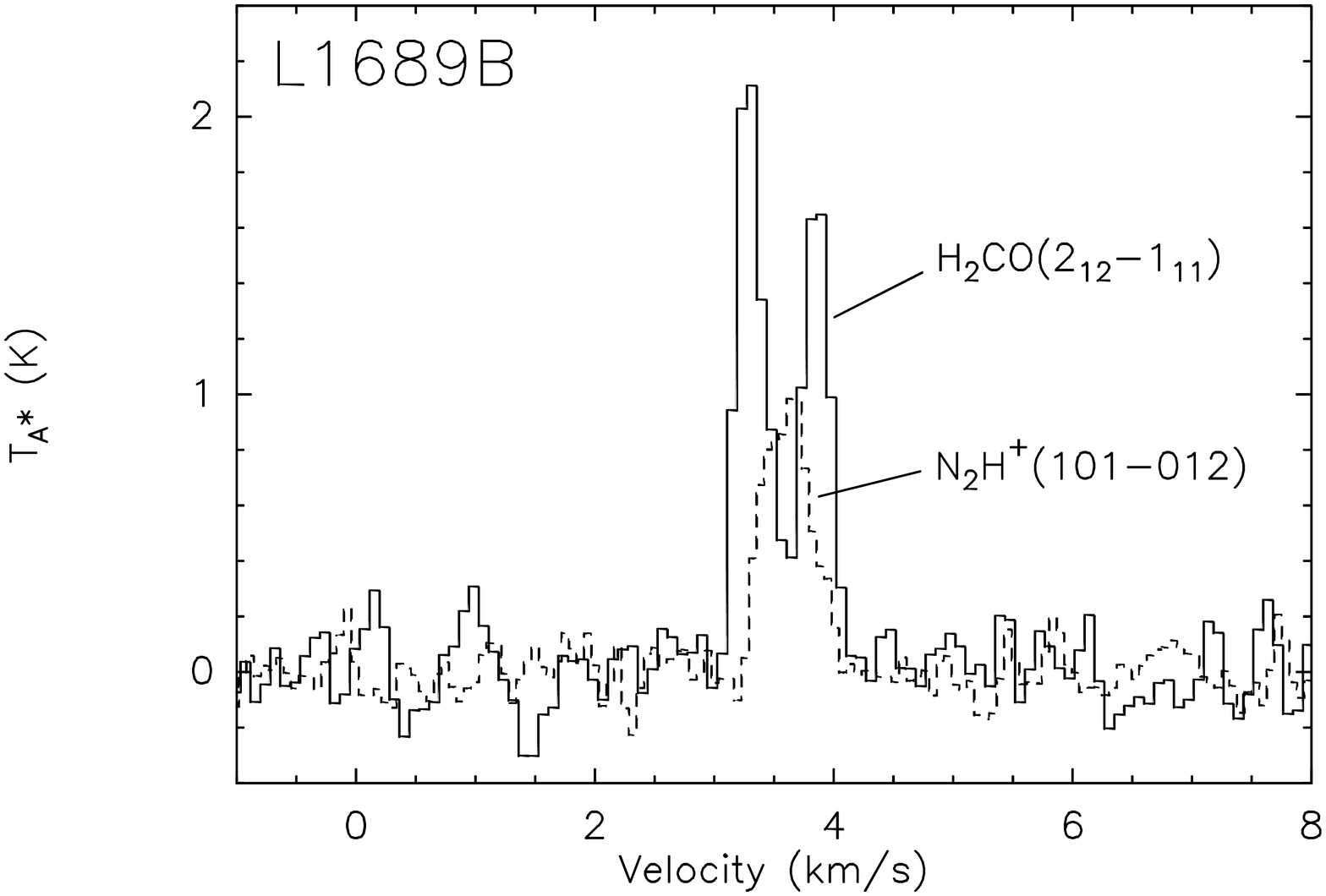,width=8cm,angle=-90}
\caption{H$_{2}$CO($2_{12}-1_{11}$) and N$_{2}$H$^{+}$(101-012) line profiles 
(solid and dashed histograms, respectively) observed by us toward the 
center of L1689B with the IRAM 30m telescope}
\label{infall}
\end{figure}
contracting is provided by the shape of the line profiles we could observe  
in optically thick molecular transitions such as H$_{2}$CO($2_{12}-1_{11}$) 
with the IRAM 30m telescope (cf. Fig.~\ref{infall}). 
Toward the center of the L1689B core, the H$_{2}$CO($2_{12}-1_{11}$) line 
profile 
is self-absorbed and skewed to the blue, while optically thin lines (such as 
the isolated hyperfine component of the N$_{2}$H$^{+}$(1-0) transition) are 
single-peaked and peak at the dip of the H$_{2}$CO self-absorption. 
This ``infall asymmetry'' is a classical indicator of significant 
inward motions (e.g. Zhou et al. 1993, Evans 1999 and references therein). 
From the shape of the H$_{2}$CO($2_{12}-1_{11}$) line profile shown 
in Fig.~\ref{infall}, an inward speed of $\sim 0.07$~km~s$^{-1}$ 
can be inferred for L1689B using the simple analytic model of 
Myers et al. (1996). A similar $\sim 0.1$~km~s$^{-1}$ infall velocity 
was reported by 
Tafalla et al.~(1998) for L1544 over spatial scales of $\sim 0.01$~pc 
to $\sim 0.1$~pc (see also Williams et al. 1999).


A variant of the magnetic models considers a non-rotating core, in which
interactions with the interstellar UV radiation 
field and between neutral and charged grains are taken into account 
(Ciolek \& Mouschovias 1994, 1995 -- hereafter CM95). 
\begin{figure*}[t]
\psfig{file=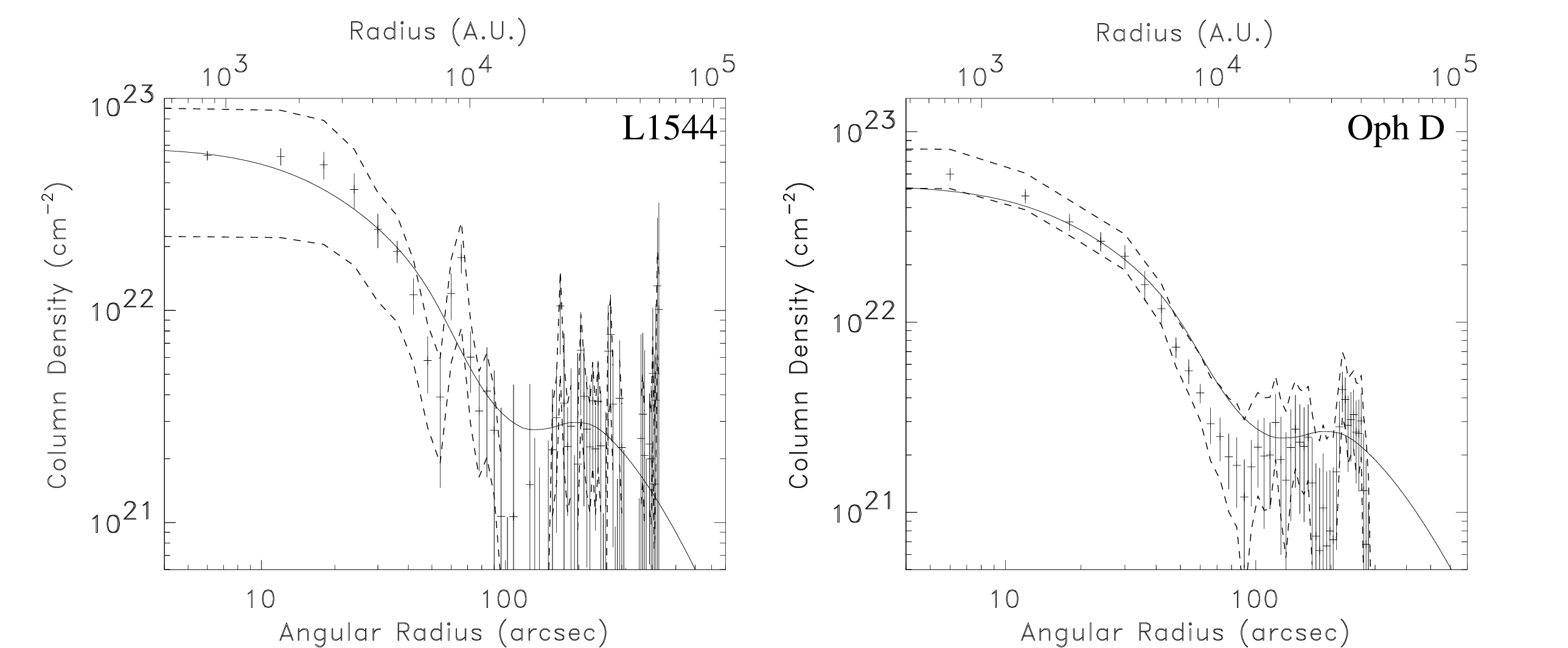,width=19cm,angle=-90}
\caption{Comparison of the magnetically supported core model B$_{UV}$ 
(at time t$_{2}$) from CM95 with the column density profiles of L1544 
and Oph D (see Table \ref{magn} for model parameters).}
\label{ciolekmodel}
\end{figure*}
The effect of the UV field is to increase the degree of ionization 
in the envelope\footnote{According to CM95 (see also McKee 1989), UV ionization exceeds cosmic ray ionization for A$_{v}<4$, which means that 
the UV field can influence cloud support up to A$_{v}\sim4$, while it excites UIB carriers only up to A$_{v}\sim$~1 (see Sect. \ref{modelling}).}, 
thereby improving the magnetic support of the outer 
enveloppe and making ambipolar diffusion less effective there. As a result, 
once a 
supercritical core has formed, the gap between the rapidly infalling central
core and the magnetically supported envelope is enhanced with respect to the
situation where cosmic rays are the only source of ionization.
We thus adopted model B$_{UV}$ of CM95 to fit the profiles of those of the 
cores that have an edge (see, e.g., Fig.~\ref{ciolekmodel}). 

Using either model 7 of BM95 or model B$_{UV}$ of CM95, we were able to find a set of
parameters yielding reasonably good fits (reduced $\chi^{2}\sim$~1--5) for most of our
cores (see Table \ref{magn}). 
We assumed the cores to be thin disks inclined by an angle $i$ 
with respect to the line of sight and applied a scaling factor $\gamma$ in column density (and $1/\gamma$ in radius) 
to the nominal dimensional models given by
BM95 and CM95, so as to bring the width of these models in agreement with that of the cores (while leaving the dimensionless parameters unchanged).

There are two problems, however, with these ambipolar diffusion models which involve only a simple static magnetic field. 
First, they produce highly flattened, disk-like structures and have 
difficulties explaining the large-scale, filamentary shape 
characterizing many of the cores we observed (e.g. Fig. \ref{absorption}). 
A disk-like cloud could appear elongated/filamentary if seen 
nearly edge-on (cf. Ciolek \& Basu 2000), but viewing most cores 
edge-on is statistically unlikely. 
The addition of turbulence in the models could perhaps alleviate 
this problem since simulations of turbulent molecular clouds tend to 
produce complex filamentary structures (e.g. Padoan et al. 1998, 
Ostriker, Gammie, \& Stone 1999). 
Although the molecular lines observed toward the core centers  
tend to be rather narrow (see, e.g., Fig.~\ref{infall} where 
$\Delta V[N_2H^+(101-012)]\sim$ 0.4 km/s at the centre of L1689B; see 
also Tafalla et al. 1998), 
indicating little turbulence in the core interiors, significant 
turbulent motions are likely to exist in the outer regions, i.e., near or
beyond $R_{edge}$ (cf. Goodman et al. 1998).

Second, and more fundamentally, they require fairly large values for the 
static magnetic field, $B_{ref}$, in the (large-scale) background 
around the cores. 
Indeed, it appears that ambipolar diffusion models can explain 
the development of 
sharp edges with $s \simgt 3$ only if they are {\it highly subcritical} 
initially (see Fig.~8 of BM95). 
For instance, model~7 of 
BM95, which provides a good fit to the profiles of L1689B (see Fig. \ref{basumodel} and Table
\ref{magn}), has an initial 
central mass-to-flux ratio $\mu_{c0} = (M/\Phi)_{c0} = 0.1$ (in units of the critical value), implying $B_{ref} \sim 170\ \mu$G. 
This strong background magnetic field is difficult to reconcile with available 
Zeeman measurements which 
indicate typical field strengths (or upper limits) 
$\simlt $~10--20~$\mu $G on $\sim$~pc scales
in low-mass star-forming clouds such as  
Ophiuchus and Taurus (e.g. Troland et al. 1996, Crutcher 1999). 
The models taking UV ionization into account (CM95) are more satisfactory 
in this respect, since they can explain some of our profiles (L1544, Oph D) 
while having a background magnetic field of only 
$B_{ref} \sim 35-50\ \mu$G (cf. Table \ref{magn}), in better agreement 
with the field values measured in dark clouds. Nevertheless, even the CM95
model is highly subcritical with an initial central mass-to-flux ratio 
$\mu_{c0} = 0.256$. Such a low value of $\mu_{c0}$ 
must be rare in actual cloud cores, since a recent analysis of 
available magnetic field observations suggests that molecular cloud cores 
are most likely supercritical by a factor of $\simlt 2$ on average 
(Crutcher 1999). Uncertainties in the measurements and in the appropriate
corrections for projection effects still allow critical or marginally 
subcritical clouds at the present time, but values of $\mu_{c0}$ 
significantly smaller than 0.5 seem unlikely. 
That cloud cores are not highly subcritical, i.e., not strongly magnetized, 
is further supported by the large values of the infall speeds measured 
in some of them such as L1544 (Williams et al. 1999), 
as well as by the slight misalignment 
between the direction of the magnetic field and the minor axis 
of the cores observed through submillimeter polarimetry 
(Ward-Thompson et al. 2000).
 
It is noteworthy that the ambipolar diffusion 
model recently proposed by Ciolek \& Basu (2000) for L1544, which uses 
$\mu_{c0} = 0.8$ and $B_{ref} \approx 12\ \mu$G (consistent with 
Zeeman measurements -- Crutcher 1999) and 
matches the central column density distribution of Fig.~5a as well as the 
$\sim 0.1$~km~s$^{-1}$ infall velocities measured by  Williams et al., 
does {\it not} account for the sharp edge ($s \simgt 4$) we 
infer in this core (cf. Table~\ref{coldens} and Fig.~1b of Ciolek \& Basu). 

To summarize, although published ambipolar diffusion models can potentially 
explain several of the features we observed, it is becoming increasingly 
clear that they do not include all the relevant effects and are not 
entirely satisfactory. 
In particular, the influence of a turbulent, non-static magnetic field 
on the (outer) structure of pre-stellar cores should be investigated more 
quantitatively 
in the future\footnote{When evaluated for the CM95 and Ciolek \& Basu (2000) models, the cutoff wavelength below which Alfv\'en waves cannot propagate in 
the neutrals (e.g. Eqs. [1a] and [2a] of Mouschovias 1991) is 
$\lambda_{A} \sim $~0.08--0.14~pc, i.e., comparable to $R_{edge}^{ISO}$ 
in Table~\ref{coldens}. This suggests that hydromagnetic turbulence may indeed play a 
role near or beyond $R_{edge}^{ISO}$ but quickly decays on smaller scales.}.

\begin{table*}
\caption[]{Parameters of adopted magnetic models}
\begin{minipage}{17.5cm}
\begin{flushleft}
\begin{tabular}{lcccccc} \hline
Source & Model & Inclination angle $i$ & Scaling factor $\gamma$ & $n_{c,0}$
(cm$^{-3}$) &
$B_{ref}$ ($\mu$G) & reduced $\chi^{2}$\\
\hline \hline
L1544 & model B$_{UV}$ of CM95 at t$_{2}$& 75~\degr & 1 & 2.6$\times$10$^{3}$ & 35 & 5.2 \\
Oph D & model B$_{UV}$ of CM95 at t$_{2}$& 75~\degr & 1 & 2.6$\times$10$^{3}$ & 35 & 1.3 \\
L1709A & model 7 of BM95 at t$_{2}$& 70~\degr & 1.5 & 4.5$\times$10$^{3}$ & 150 & 3 \\
L1689B  & model 7 of BM95 at t$_{2}$& 65~\degr & 1.7 & 5.1$\times$10$^{3}$ & 170 & $\sim$~2\\
L328 & model B$_{UV}$ of CM95 at t$_{2}$& 0~\degr & 1.4 & 3.6$\times$10$^{3}$ & 50 & 6.8\\
L429 & model 7 of BM95 at t$_{3}$& 65~\degr & 1 & 3$\times$10$^{3}$ & 100 & 5.7 \\
GF5 & model 7 of BM95 at t$_{2}$& 60~\degr & 1 & 3$\times$10$^{3}$ & 100 &3.3 \\
\hline 
\end{tabular}
Notes:\\
(1) $t_2$ (resp. $t_3$) is the time at which the central number density has increased
by a factor of 10$^{2}$ (resp. 10$^{3}$)\\
(2) $n_{c,0}$ is the initial central density\\
(3) $B_{ref}$ is the large-scale magnetic field in the reference state\\
(4) $\mu_{c0}= (M/\Phi)_{c0} = 0.1 (M/\Phi)_{crit}^{cent}$ for model 7 of BM95 and 
(5) $\mu_{c0}=0.256 (M/\Phi)_{crit}^{cent}$ for model B$_{UV}$ of CM95\\
(6) $\gamma$ is the scaling factor by which we have to multiply the column density and to divide the radius of the models given by BM95 and CM95 so as to adjust the widths of the models to that of our cores.\\
(7) The temperature of the cloud is 10~K.\\
\end{flushleft}
\end{minipage}
\label{magn}
\end{table*}

\subsection{Filamentary structures and cloud fragmentation} 

As already stated (\S \ref{BE}), most of the cores observed
here depart from the spherical morphology adopted in many theoretical models
(e.g. \S \ref{BE} to \ref{singular}). In particular, GF5 has an elongated, filamentary aspect
reminiscent of L977 and IC5146 discussed by Alves et al. (1998) and Lada, Alves
\& Lada (1999). These authors use extinction measurements of background
starlight in the near-IR to derive the density structure of their cores. 
They model L977 and IC5146 as cylindrically symmetric cores with 
$\rho \propto r^{-2}$ density gradients, and point out that their 
data are {\it not} consistent with models of 
isothermal cylinders (which predict $\rho \sim r^{-4}$ -- Ostriker 1964). 
Lada et al. thus invoke new models of cylindrical clouds threaded by helical magnetic fields (Fiege \& Pudritz 1999a), which yield density profiles 
falling off as $r^{-1.8}$ to $r^{-2}$ in agreement with their extinction 
results. Our data however (see, e.g., GF5 in Fig. \ref{absorption}l and Fig. \ref{isomm}h) 
suggest that such elongated cores are made up of a succession of discrete  spheroidal condensations with outer density gradients approaching 
$\rho \propto r^{-2}$, rather than of a single continuous, 
cylindrical filament with $\rho \sim r^{-2}$.
Our interpretation is supported by the fact that the condensations of, e.g., 
GF~5 and Oph D are seen in more than one tracer or wavelength.

We have derived the (half-power) radii ($R$~$\sim$~3500--12000~AU), mean column densities ($<N_{H_{2}}>$$\sim$ 
2--6$\times 10^{22}$~cm$^{-2}$) and densities
($<n_{H_{2}}>\sim$10$^{5}$--10$^{6}$~cm$^{-3}$) of the three condensations of the GF5 filament and
of the two condensations of the Oph D core and inferred their masses from our ISOCAM
and 1.3~mm continuum measurements. The estimated condensation masses, around 0.9 M$_{\odot}$ for GF5
and 0.3 M$_{\odot}$ for Oph D are consistent with, or slightly larger than, the Bonnor-Ebert critical mass
(e.g. Bonnor 1956), $M_{BE}=2.4 R a_{s}^{2}/G$ (where $a_{s}$ is the isothermal
sound speed), of 0.7 M$_{\odot}$ and 0.2 M$_{\odot}$, respectively, for a gas
temperature of 10~K. This suggests that the condensations are
self-gravitating ($M\geq M_{BE}$). In the case of Oph D for which we have DCO$^{+}$(2-1)
observations, we also 
calculated the virial masses, $M_{vir}=5R\sigma_{tot}^{2}/G$, of the
condensations assuming their density is uniform (which is consistent with
the flattening of the column density profile of the cores in their inner part).
The measured DCO$^{+}$(2-1) linewidths are $\sim$0.5~km~s$^{-1}$, yielding $M/M_{vir}\sim$~0.5 and
confirming that these structures are gravitationally bound (cf. Pound \& Blitz
1993). The characteristics of the GF5 and Oph D condensations are reminiscent 
of those found for the pre-stellar fragments of the $\rho$ Oph protocluster (MAN98). The
average spacing between two condensations is 20000~AU for
GF5 and 9000~AU for Oph D, which is of the order of the Jeans diameter
$\lambda_{J}\simeq 12000 $~AU in the parent cloud core, for a gas temperature 
of 10~K and an average
density of $10^{5}$~cm$^{-3}$. Furthermore, the ratio of condensation spacing 
to filament diameter is $\sim$~2, consistent with models where gravity plays a major role in fragmenting the cores/filaments (Larson 1985, Nakamura et
al. 1993, Fiege \& Pudritz 1999b).

The existence of these compact condensations also emphasizes the fact that 
even isolated prestellar cores (such as most cores in our sample) 
are fragmented and are likely to form more than one star, perhaps even small
groups of stars (cf. Tafalla et al. 1999).

\subsection{Implications for the earliest stages of collapse}

The presence of flat inner radial density gradients and sharp outer edges 
in the pre-stellar cores of our sample 
suggests that the initial conditions for fast 
protostellar collapse are often neither singular nor scale-free. 
In many cases, the collapse initial conditions probably differ 
substantially from the idealized singular isothermal sphere (SIS) adopted by 
the standard theoretical scenario of Shu et al. (1987) which leads to 
a constant mass accretion rate during the protostellar phase. 
It is likely that  protostar evolution will depart, at least 
initially, from the predictions of the standard paradigm, 
even if this model provides a good, first-order approximation 
of self-initiated, isolated star formation. 
When starting from Bonnor-Ebert-like initial conditions, 
all (magneto-)hydrodynamic collapse models 
predict a {\it decline of the accretion rate} 
during the protostellar phase (e.g. Foster \& Chevalier 1993, 
Tomisaka 1996, Henriksen et al. 1997, 
Basu 1997, Safier, McKee, \& Stahler 1997, Ciolek \& K\"onigl 1998, Li 1999). 
Interestingly enough, observational evidence for a declining protostellar 
accretion rate has been found in the form of a marked decrease 
of outflow power\footnote{Since protostellar outflows are believed 
to be directly powered by accretion (e.g. K\"onigl \& Pudritz 2000), 
the outflow momentum rate may be used as a surrogate tracer of the 
accretion rate.} with time from young (Class 0) protostars 
to evolved (Class I) protostars (Bontemps et al. 1996).

Another related implication of the sharp outer edges we observed is that
most pre-stellar cores appear to be effectively decoupled from their 
large-scale parent clouds, providing a finite reservoir of mass for 
subsequent star formation. 
This supports the view according to which there is a mass scale in the 
star formation process (e.g. Larson 1999) 
and the stellar initial mass function is at least 
partly determined by fragmentation at the pre-collapse stage (see also 
MAN98).

\section{Summary and conclusions}

We imaged a sample of 24 pre-stellar cloud cores at 7~$\mu$m with 
the mid-IR camera ISOCAM aboard the ISO satellite.  
Our main results and conclusions are the following:

\begin{enumerate}
\item As many as 23 of the 24 fields of our programme show absorption-like 
features in the mid-IR. By observing and detecting 8 of these in 
{\it emission} at 1.3~mm with the IRAM~30~m telescope, we show 
that most of the mid-IR dark features trace dense, cold core material seen 
in absorption against a diffuse background emission arising from the rear 
side of the parent molecular cloud.

\item Based on a simple analysis of the mid-IR intensity profiles, 
and using independent constraints on the central and outer column densities 
from millimeter continuum and line data, we are able to derive the mean 
radial column density profile of each core detected in absorption with ISOCAM.

\item We confirm that pre-stellar cores have density and column density 
profiles that flatten out near their centers, a conclusion already reached 
by Ward-Thompson et al. (1994, 1999) and Andr\'e et al. (1996) on the basis of (sub)millimeter emission 
maps. At radii less than $R_{flat} \sim $~4000--8000~AU 
(i.e. $\sim $~0.02--0.04~pc) the derived radial profiles are significantly 
flatter than $\rho \propto r^{-2}$ and $N_{H_{2}} \propto \bar{r}^{-1}$
(i.e., as flat as $N_{H_{2}} \propto \bar{r}^{-0.15}$--$\bar{r}^{-0.2}$), respectively.

\item We find that, beyond $R_{edge} \sim $~0.05--0.3~pc, 
at least three (and possibly four) of the cores in our sample are 
characterized by sharp edges corresponding to radial column density 
profiles that drop  
steeper than 
$N_{H_{2}} \propto \bar{r}^{-2}$ with projected radius. Depending on the details 
of deprojection effects, this suggests that the {\it density gradient} becomes 
itself steeper than $\rho \propto r^{-3}$ near $R_{edge}$.
An important implication is that the mass reservoir available for star 
formation in these cores is finite, supporting the idea that stellar 
masses are partly determined at the pre-stellar stage.

\item Between $R_{flat}$ and $R_{edge}$, the core radial profiles 
approach a near power-law regime consistent with $N_{H_{2}} \propto \bar{r}^{-1}$
 and $\rho \propto r^{-2}$. In particular, we show that models 
predicting much flatter power-law radial density profiles (e.g. $\rho \propto r^{-1}$) 
such as logotropic non-isothermal spheres (e.g. McLaughlin \& Pudritz 1996) 
cannot account for the structure of our low-mass pre-stellar cores.

\item Points (3) and (4) above suggest that the initial conditions for fast 
protostellar collapse are neither singular nor scale-free. Semi-analytical 
collapse models based on self-similar initial conditions such as the 
singular isothermal sphere or the singular logotrope 
can thus provide only 
a first-order, approximate description of the protostellar phase.

\item We have compared our results with various theoretical models of core
structure, and have found that ambipolar diffusion models of 
magnetically-supported cores (e.g. Ciolek \& Mouschovias 1995, 
Basu \& Mouschovias 1995) are 
most promising although they require a (probably unrealistically) strong 
background magnetic field ($\sim$~30--100~$\mu$G). We hypothesize that 
more elaborate versions of these models, incorporating the effects 
of a non-static, turbulent magnetic field in the outer parts of the cores
would be more satisfactory and could also account for the filamentary 
shape often seen on large ($\simgt 0.25$~pc) scales.
\end{enumerate}

\bigskip
{\it Acknowledgements.} We would like to thank Glenn Ciolek and 
Shantanu Basu for enlightening discussions on ambipolar diffusion and 
for kindly providing us with the magnetic models shown in Figs. \ref{basumodel} 
and \ref{ciolekmodel}. We are also grateful to Bill Reach for giving us estimates of the Zodiacal light emission based on modeling of the 
DIRBE/COBE data, and to Malcolm Walmsley and Patrick Boiss\'e for their 
useful comments.

\end{document}